\documentclass[journal,review,submit,pdftex,oneauthors]{mdpi} 

\usepackage{tikz}

\firstpage{1} 
\pubvolume{1}
\issuenum{1}
\articlenumber{0}
\pubyear{2025}
\copyrightyear{2025}
\datereceived{ } 
\daterevised{ } 
\dateaccepted{ } 
\datepublished{ } 
\hreflink{https://doi.org/} 

\newcommand{\beq}{\begin{equation}}
\newcommand{\eeq}{\end{equation}}

\def\asec{\ifmmode ^{\prime\prime}\else$^{\prime\prime}$\fi}

\def\Ms{\mbox{\,M$_{\odot}$}}

\def\msun{\hbox{M$_{\odot}$}}

\def\msunyr{\mbox{\,${\rm M_{\odot}\, yr^{-1}}$}}
\def\mdot{\dot M}
\def\Mdot{\dot M}

\def\degs{\ifmmode ^{\circ}\else$^{\circ}$\fi}
\def\amin{\ifmmode ^{\prime}\else$^{\prime}$\fi}
\def\asec{\ifmmode ^{\prime\prime}\else$^{\prime\prime}$\fi}

\def\degs{\ifmmode ^{\circ}\else$^{\circ}$\fi}
\def\amin{\ifmmode ^{\prime}\else$^{\prime}$\fi}

\def\cm{\mbox{\,cm}}

\def\cm3{\mbox{\,cm$^{-3}$}}
\def\kms{\mbox{\,km~s$^{-1}$}}

\def\kms{\mbox{\,km s$^{-1}$}}

\def\lsim{\!\!\!\phantom{\le}\smash{\buildrel{}\over
 {\lower2.5dd\hbox{$\buildrel{\lower2dd\hbox{$\displaystyle<$}}\over
                                 \sim$}}}\,\,}
\def\gsim{\!\!\!\phantom{\ge}\smash{\buildrel{}\over
{\lower2.5dd\hbox{$\buildrel{\lower2dd\hbox{$\displaystyle>$}}\over
                               \sim$}}}\,\,}

\Title{Radio Supernovae}

\TitleCitation{Radio Supernovae}

\Author{Esha Kundu$^{1}$\orcidA{}}

\AuthorNames{Esha Kundu}

\isAPAStyle{%
       \AuthorCitation{Kundu, E.}
         }{%
        \isChicagoStyle{%
        \AuthorCitation{Kundu, Esha}
        }{
        \AuthorCitation{Kundu, E.}
        }
}

\address{%
$^{1}$ \quad Department of Physics, Indian Institute of Technology (Indian School of Mines) Dhanbad, Dhanbad -826004, Jharkhand, India; eshakundu@iitism.ac.in}
\corres{Correspondence: eshakundu@iitism.ac.in}

\abstract{Supernovae (SNe), the catastrophic end of stars' lives, are among the most energetic phenomena in the universe. Mapping the aftermath of the explosions to the properties of pre-SN stars is challenging due to the lack of knowledge about the evolution of different types of stars. The immediate surroundings of pre-SN stars carry the signature of the progenitors, and radio observations are the best way to examine the ambient media. Since radio emission originates from the interaction of supersonic SN ejecta with the relatively stationary circumstellar medium, with a few years of radio study, the mass-loss history of progenitor stars can be probed from just before the explosion of the star to thousands of years before the onset of the SN. Moreover, this can provide crucial details about the explosions, which are poorly understood to date. In this paper, we review the radio properties of different types of core-collapse explosions and thermonuclear runaways to understand their mass-loss evolution---which allows us to unravel the imprints of the progenitors on the surrounding media and thus the nature of the exploded stars. Additionally, we discuss the current state of the art in this field, including existing and the next-generation radio facilities with enhanced capabilities that provide further details about these explosions.}


\begin{document}




\section{Introduction}
\label{intro}

Radio emission from a supernova (SN) mainly originates from the interaction of supersonic SN ejecta with the circumstellar medium (CSM). This ambient medium is created by the mass loss from the progenitor star in different evolutionary phases. In~the case of a core-collapse SN, mass loss toward the end of the star's life affects its final phases significantly and thus influences the core-collapse process (e.g., see \citep{SA14, smith14}). 
For thermonuclear explosions, similarly, mass loss  plays a crucial role in the evolution of their progenitors (e.g., see thermonuclear runaways of white dwarfs (WDs) through different channels \citep{HF90,WI73, IT84, webbink84}). 


\par 
Almost all types of core-collapse SNe have been detected at radio wavelengths. In~contrast, only one Type Ia explosion, SN~2020eyj, has been detected at this frequency \citep{kool23} after decades of observational campaigns in detecting these objects at radio waves. However, among~all core-collapse SNe, only around 30\% of these events were bright enough at radio wavelengths to be detected by radio facilities \citep{Bietenholz21a}. Even a significant number of nearby explosions were never detected at radio wavelengths. The~light curves of detected SNe show significant variations with age and subtypes. However, because~of an insufficient statistical sample, it becomes difficult to parameterize the radio light curves of different subtypes of SNe (nevertheless, see \citep{weiler02, laura16,Bietenholz21a}, where attempts were made to parameterize radio light curves of subclasses of core-collapse and thermonuclear SNe). 
One of the brightest SNe at radio wavelength in terms of the peak spectral luminosity ($L_{\nu}^{pk}$) is SN 1998bw, with $L_{\nu}^{pk} \sim 10^{29}$~erg/s/Hz \citep{kulkarni98}, which is associated with $\gamma$-ray burst GRB980425. 
Nevertheless, a~considerable fraction of SNe have $L_{\nu}^{pk}$ much lower than that of SN 1998bw. In~addition, the~time to reach the peak spectral luminosity differs from event to event. For~progenitors with a significant mass loss just before the collapse, which is often the case for Type IIn explosions, the~peak spectral luminosity is reached at a significantly later stage. For~example, for~SN 1986J, it takes around a year to attain the peak, while for~SN 1987A, which is a Type IIP explosion, the~peak occurs within a day \citep{turtle87}. However, increased radio emission was observed from this SN at a later time. Explosions such as SN 2014C, which was initially classified as a Type Ib explosion, later showed enhanced radio emission due to interaction with a large amount of ambient material, which was expelled from the progenitor sometime before the explosion \citep{margutti17,bietenholz18}. One of the best-studied SNe at radio wavelength is Type IIb explosion SN 1993J.
The light curves of this event displayed a sudden dip around 3000 days after the explosion, which was most likely due to a sudden decrease in mass-loss rates of the progenitor around 7000 years prior to its explosion \citep{weiler07, kundu19}.

In the case of Type Ia explosions,~single radio-detected event SN~2020eyj allows for detection at radio wavelengths (at 5.1 GHz) more than approximately 1.5 years after the explosion. Recently, with~the advancement of current-generation radio telescopes, by~using them in wide-field radio surveys, such as the Very Large Array (VLA; \citep{Thompson80,Napier83,Perley11}) Sky Survey (VLASS), and~variable and slow transients (VAST) Pilot surveys \citep{Murphy21} utilizing the Australian Square Kilometre Array Pathfinder (ASKAP; \citep{Hotan21}), it has become feasible to reconstruct the mass-loss history of progenitor stars before their death by detecting the rebrightening of these explosions at late times, which provides robust constraints on the evolution of pre-SN~stars.

\par 
In this paper, we review the radio properties of different types of core-collapse and thermonuclear runaways to understand their mass-loss evolution,~hence the progenitors of these explosions, along with the current state of the art in this field. Detailed reviews of radio emission from supernovae until around the previous decade were summarized by Weiler~et~al. (2002, 2010) \citep{weiler02,weiler10}.  
The layout of this paper is as follows: In the following section, Section 
\ref{sn_csm_int}, the~interaction of SN ejecta with CSM is discussed. 
In Section~\ref{rad_emissn},  the~theory of radio emission from SNe is described. The~observations of radio emission from different types of core-collapse and thermonuclear explosions are discussed in Section~\ref{rad_sne}. Finally, the~discussion and conclusions are given in Sections~\ref{discusn} and \ref{conclu}, respectively.

\section{SN-CSM~Interaction}
\label{sn_csm_int}
\begin{figure}[H]
\includegraphics[width=0.53\textwidth]{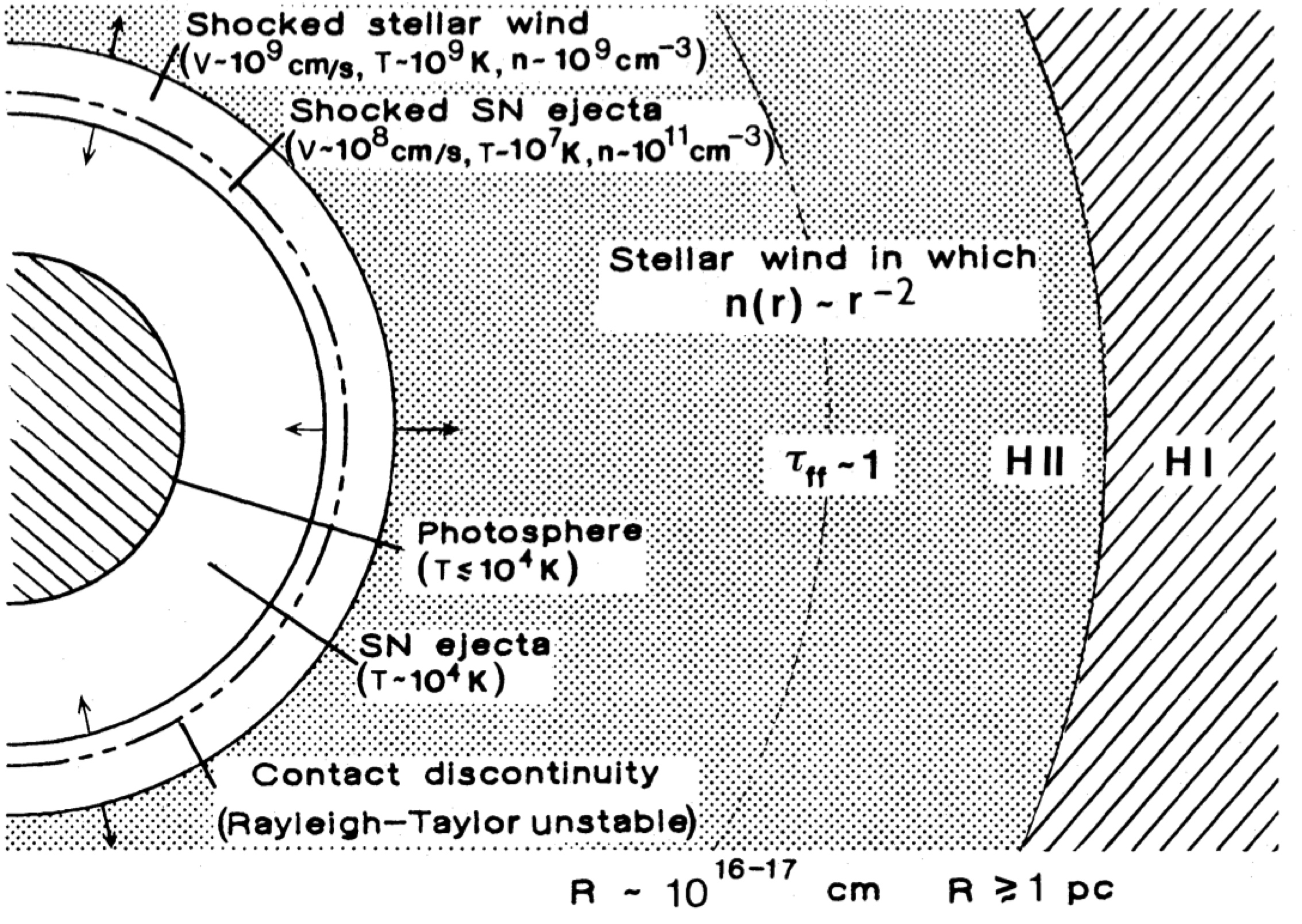}
\includegraphics[width=0.38\textwidth]{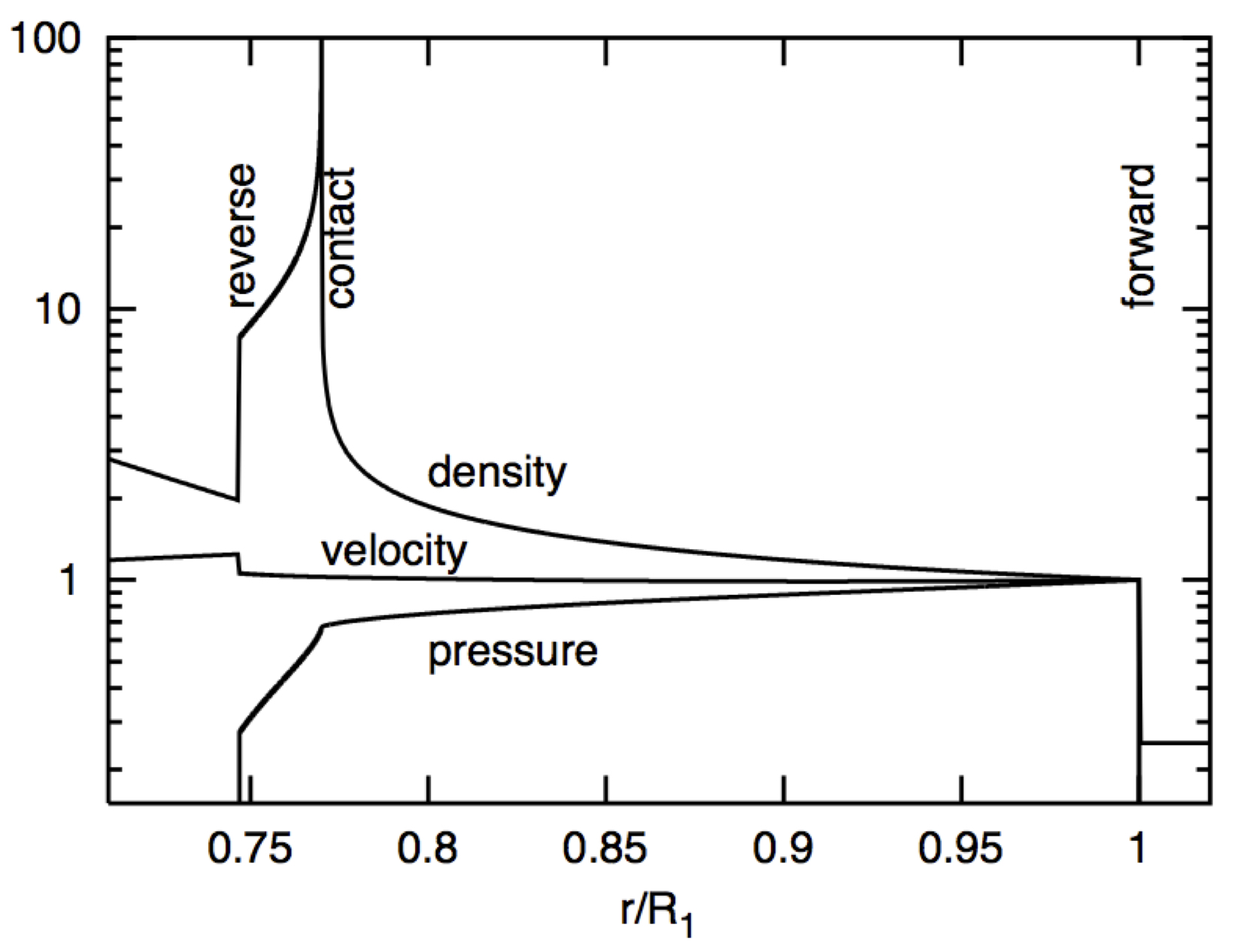}
\caption{(\textbf{{Left}
} panel): Interaction of SN ejecta with a wind-like ambient medium. The~different regions shown here are not to-scale. The~contact discontinuity exists between the forward and reverse shocks. Figure reproduced from  \citep{peter88} with permission. (\textbf{Right} panel): For a wind-like CSM and ejecta with an extreme outer part having a power-law profile with a power-law index of 7, the~density, velocity, and~pressure profiles are displayed within the shocked shell, which is bounded by the forward and reverse shocks. The~contact discontinuity between~the two shocks is also depicted here. The~X-axis is normalized with respect to the position of the forward shock (which is represented here by $\rm R_1$), and~the Y-axis is scaled according to pressure, velocity, and~density values just behind the forward shock. Figure reproduced from  \citep{blondin01} with~permission.}  
\label{SN_CSM_int}
\end{figure}

\begin{figure}[H]
\includegraphics[width=0.9\textwidth]{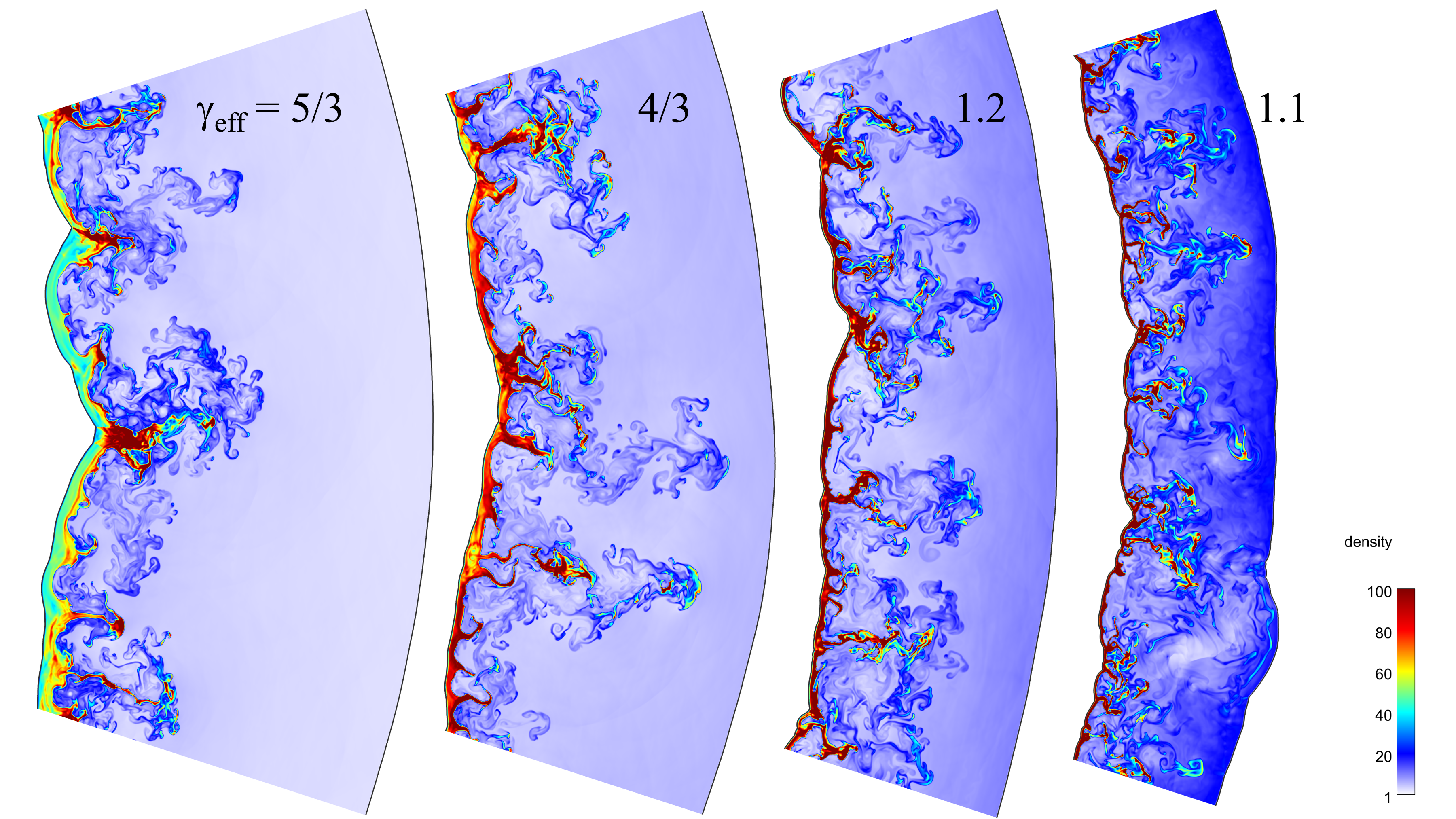}
\caption{The growth of the RT instability in the shocked ejecta for different effective adiabatic indices, $\gamma_{\rm eff}$, of~the gas. This simulation is performed for a wind-like CSM and ejecta with an extreme outer part having a power-law profile with a power-law index of 7. The~finger-like structures represent the RT instability with the density color bar scaled with respect to the density of the CSM just ahead of the forward shock. For~higher $\gamma_{\rm eff}$, the~instability stays close to the reverse shock area, while as $\gamma_{\rm eff}$ decreases, the~protrusions spread across the entire shocked region. Figure reproduced from  \citep{kundu_thesis_2019} with~permission.}
\label{RT_insbility}
\end{figure}

After the explosions of stars, either in the core-collapse or thermonuclear scenario, the~SN ejecta are thrown into the CSM with a speed $ \geq$$10^{4}$ \kms~\citep{chevalier76}. The~velocity of the circumstellar medium is usually significantly lower compared with the velocity of the ejecta. As~a result, the~interaction of this high-velocity ejecta with the almost stationary ambient medium launches two shocks, with the first being the forward shock, which plows into the CSM, hence called the forward shock. The~enormous pressure behind the forward shock launches another shock, which moves backward in the mass coordinates. This shock is called the reverse shock. The~region between these two shocks contains shocked CS and ejecta material, as well as~a contact discontinuity, as~demonstrated in the left panel of Figure~\ref{SN_CSM_int} for a wind-like ambient medium (figure reproduced from \cite{peter88}). The~variations in density, velocity, and~pressure in this shell as a function of radius are displayed in the right panel of Figure~\ref{SN_CSM_int} (figure reproduced from \cite{blondin01}). The~region between the reverse shock and contact discontinuity contains shocked ejecta, which induce Rayleigh--Taylor (RT) instability \citep{chevalier82b}. 
 This instability arises likely due to the presence of low-density material (see the right panel of Figure~\ref{SN_CSM_int}) behind~the forward shock, which affects the expansion of the high-density matter behind the reverse shock. The~growth of RT instability in the shocked ejecta for different effective adiabatic indices ($\gamma_{\rm eff}$) of the gas is shown in Figure~\ref{RT_insbility}. This simulation is performed for a wind-like CSM and ejecta with an extreme outer part having a power-law profile with a power-law index of 7 (see Section~\ref{sn_ejecta} for details about the ejecta profile). The~finger-like structures represent the RT instability, with the density color bar scaled with respect to the density of the CSM just ahead of the forward shock. As~demonstrated, for~higher $\gamma_{\rm eff}$, the~instability stays close to the reverse shock area, while as $\gamma_{\rm eff}$ decreases, the~protrusions spread across the entire shocked region \citep{kundu_thesis_2019}.

\unskip

The shocked region between the reverse and forward shocks is an ideal place for particle acceleration and magnetic field amplification \citep{bykov13, caprioli14a, caprioli14b}. Relativistic electrons gyrating in this magnetic field produce radio emission mainly through synchrotron radiation \citep{rybicki79}. As~the flux of this radiation is roughly proportional to the density of the ambient medium, the~study of radio emission is an excellent tool to probe the mass-loss history of the progenitor star.
Moreover, since different stars lose material from their outer surfaces at different rates, this investigation allows one to gain information about the pre-SN star. The~ambient medium expected to prevail around the pre-SN star and the profile of the SN ejecta, which play crucial roles in the production of radio emissions from SNe, are discussed in Sections~\ref{csm} and \ref{sn_ejecta}, respectively.


\subsection{CSM~Structure}
\label{csm}

Massive stars, which end their life in core-collapse explosions, lose their envelope masses mainly through winds \citep{nugis00,vanloon05,mokiem07}. As~a result, the~density of the ambient medium decreases with the square of the radius. For~a constant mass-loss rate, $\dot M$, the~CSM density at a distance $r$ from the star is $\rho_{CSM} (r) = \dot M/(4 \pi r^2 v_w)$, for~a constant $v_w$, where $v_w$ represents the speed at which mass is lost from the system. For~example, compact stars like blue supergiants and Wolf--Rayet stars have wind speeds in the range of 1000~\kms to 3000~\kms, with $\dot M$ between $10^{-6}$ and 
$10^{-4} \msunyr$ \citep{Cappa04,Crowther07,smith14}. In~the case of red supergiants, $v_w$ is significantly low, between~10 and 50 \kms, and~the mass loss is around $10^{-6}${--}$10^{-5} \msunyr$. However, there exist some stars that go through enormous mass loss, with $\Ms$ $ \sim 10^{-3} \msunyr$, for a wind speed similar to that of red supergiants \citep{smith14}. 

\par 
In the case of a single degenerate (SD) explosion channel of the thermonuclear explosion, 
 similar wind-like CSM expected to prevail. 
However, the~double degenerate (DD) scenario predicts an ambient medium of constant density, which can be expressed as $\rho_{CSM} = n_{ISM} \mu$, where $n_{ISM}$ and $\mu$ represent the constant particle density and mean atomic weight of the ambient material, respectively. Therefore, we write the density of the CSM as $\rho_{CSM} = A r^{-s}$, where $s =$ 2 with $A = \dot M/(4 \pi v_w) $ for a wind-like medium and~$s =$ 0 with $A =  n_{ISM} \mu$ for constant-density medium, which is a characteristic of the interstellar~medium. 


\subsection{SN Ejecta~Profile}
\label{sn_ejecta}
\begin{figure}[H]
\includegraphics[width=0.95\textwidth]{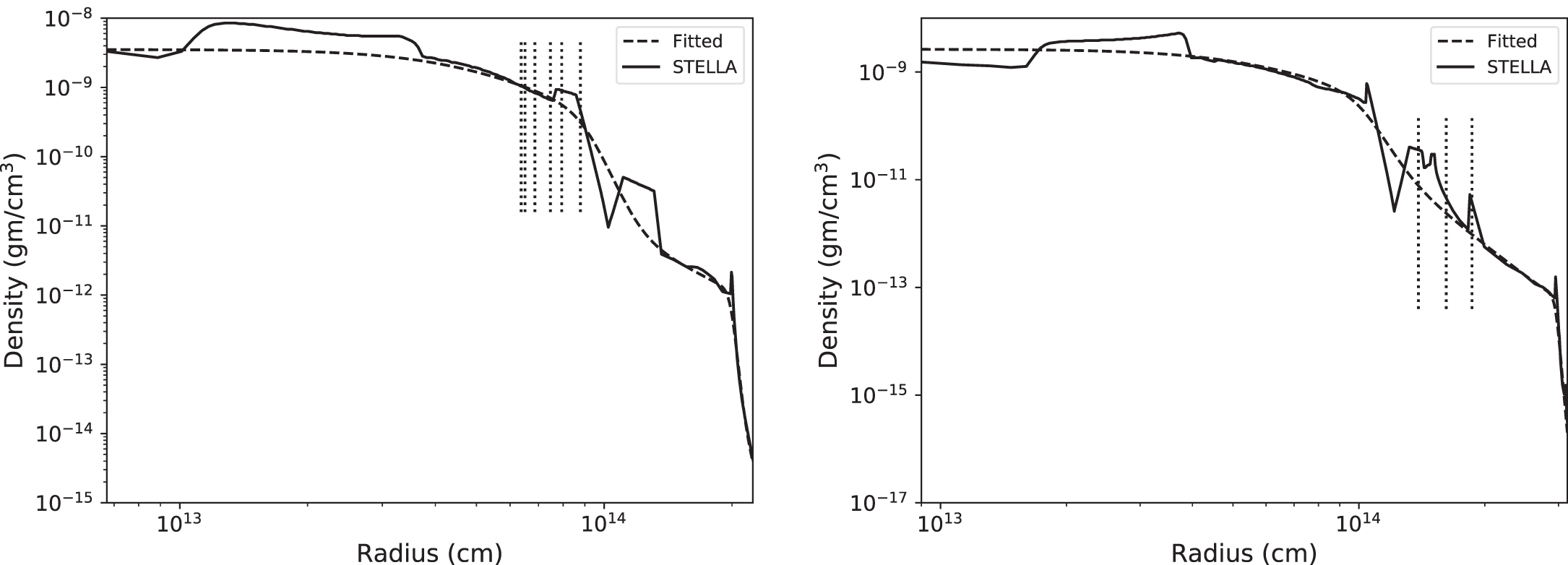}
\caption{SN ejecta profiles, as~obtained from STELLA along with the fitted ones, for~SN 1993J (\textbf{left} panel) and SN 2011dh (\textbf{right} panel), one day after the explosions. The~dotted vertical lines in the ejecta profiles illustrate the reverse shock position at 1000, 2000, 3000, 5000, 7000, and 8000 days after the explosion of SN 1993J and that at 200, 500, and 1200 days after the explosion of SN 2011dh. Figure reproduced from  \citep{kundu19} with~permission.}
\label{den_prof}
\end{figure}

Besides the density profile of the CSM, the~ejecta structure of an SN plays a crucial role in determining the flux of radio emission from the interaction region. Analytical and numerical simulations have suggested that the inner part of SN ejecta follows almost a constant density up to a given radius, $r_{brk}$, which corresponds to the break velocity, and~beyond $r_{brk}$, the density of the outer part of the SN decreases rapidly with the radius as $\rho_{SN} \propto r^{-n}$, where $n$ is the power-law index \citep{Matzner99, kundu17}. The~density structure of an SN could be even more complicated, as~demonstrated in Figure~\ref{den_prof}, where the solid lines illustrate the ejecta profiles of SN 1993J (left panel) and SN 2011dh (right panel) as obtained from numerical simulations that used multigroup radiation hydrodynamic simulations (STELLA)~\citep{kundu19}.

The interaction between SN ejecta, which have a power-law profile, and~a CSM, characterized by $\rho_{CSM} = A r^{-s}$, can be described by a self-similar structure \citep{chevalier82a}. Soon after the explosion, the~free or homologous expansion of the ejecta is established. 
The SN remains in the homologous phase as long as the reverse shock plows through the extreme outer part of the ejecta.
The duration of the free expansion phase is estimated as $t_{\rm FE} = \Lambda^{\frac{1}{(3-s)}}~ \xi_2^{\frac{n-s}{3-s}} ~ v_{o,ej}^{\frac{s}{3-s}}$, with $\Lambda = \frac{3-s}{n-3} ~ \frac{4\pi}{\theta} ~ \frac{\rho_{o,ej}}{\beta} ~ \frac{\xi_2^{3-n}}{\xi_1^{3-s}} ~ r_{o,w}^{2-s}$ \citep{kundu17}. Here, $v_{o,ej}$ and~$\rho_{o,ej}$ represent the velocity and density of the extreme outer part of the inner ejecta, respectively. $\xi_1 = r_s/r_c$, and  $\xi_1 = r_{rev}/r_c$, with $r_s$, $r_c$, and~$r_{rev}$ being the forward shock, contact discontinuity, and~reverse shock radii, respectively. For~the wind-like medium $\beta = \dot M/v_w$ and~in the case of CSM of constant density, $\beta = n_{ISM}$. $\theta$ is the ratio between the swept-up ejecta and the swept-up ambient matter, and~$r_{o,w}$ represents a reference radius. At~the end of the free expansion phase, when the reverse shock enters the flat inner part of the SN, the~explosion moves into the Sedov--Taylor (ST) phase. 

\par
For a self-similar structure, the~evolution of the radius of the forward shock ($r_s$) in the free expansion phase is $r_s \propto A^{1/(s-n) ~ t^{(n-3)/(n-s)}}$, and~the corresponding shock speed is given by $v_s = \frac{dr_s}{dt} = \frac{(n-3)}{n-s} r_s/t$, where $t$ represents time \citep{chevalier82a}. Depending on the nature of the CSM and the values of $n$, the~shocked shell can have variations in its thickness, with~the maximum thickness of the interaction region being around 30\% of the forward shock~radius.

\section{Radio Emission from~SNe}
\label{rad_emissn}

According to diffusive shock acceleration theory (DSA), in~the shocked region, the~back-and-forth movements of charged particles across the shocks accelerate them to relativistic energies \citep{bell78}. For~a compression factor of 4, DSA suggests that the accelerated particles follow a power-law energy spectrum with $dN/dE = N_0 E^{-p}$, where $N_0$ and $p$ are the normalization constant and the power-law index, respectively. However, along with the main shock, if~a sub-shock is present, the~spectrum steepens for low-energy particles with $p~>~2$~\citep{ellison91}. In~addition, when different losses occur, such as synchrotron and inverse Compton losses, or~particles escape from the shocked region, the~resulting spectrum becomes softer by another $E^{-1}$ factor. For supernovae that are believed to originate from compact progenitors, such as Type Ib/c SNe, the~power-law index of relativistic electrons is estimated to be 3 \citep{chevalier98,chevalier06}. However, for~these events, the~loss processes were found to be irrelevant. Particle acceleration is a complicated mechanism. In~addition, a~poor understanding of the acceleration process provides inadequate information on the precise value of $p$. 

\par
In the shocked region, the~electrons over the protons are mainly responsible for synchrotron radiation, since the power of this radiation decreases inversely with the mass of the charged particle \citep{rybicki79}. It is usually assumed that in~the post-shock regime, all electrons get accelerated. If~the fraction of post-shock energy density that goes to electrons is $\epsilon_e$, where $\epsilon_e = u_e/u_{th}$, for~accelerated electrons, this implies that the population has a minimum energy, $E_{min} \propto \epsilon_e ~ r_s^2 ~ t^{-2}$, for a constant value of $\epsilon_e$ \citep{kundu17}. Here, $u_{th} = \frac{9}{8} \rho_{CSM}(r) v_s^2$ is the post-shock kinetic energy density, and~$u_e$ represents the energy density of electrons. In~the initial phase, the~shock speed is so high that almost all electrons are in the relativistic regime, i.e.,~$E > m_e c^2$, where $m_e$ and $c$ are the mass of an electron and the speed of light in vacuum, respectively. As~a result, $\epsilon_e = \epsilon_{\rm rel}$, with~$\epsilon_{\rm rel} > 0.16 (v_s/5000 ~ { \kms})^{-2}$ for $p=3$~\citep{chevalier06}. However, with~time, as~the shocks get decelerated, a~non-relativistic component of the electrons starts to appear, and~we get $\epsilon_e = \epsilon_{\rm rel} + \epsilon_{\rm nrel}$, with~$\epsilon_{\rm nrel}$ representing the fraction of energy in non-relativistic electrons. As~a result, we get $\epsilon_{\rm rel} = \epsilon_e \big(\frac{E_{min}}{m_e c^2} \big)^{(p-2)}$ \citep{kundu17}. Since synchrotron emission originates from relativistic electrons, it is assumed that electrons with energies greater than $m_ec^2$ are responsible for this emission. Multidimensional hybrid simulations of non-relativistic shocks showed that energetic electrons can acquire a maximum of 10--20\% of the bulk kinetic energy of the shock \citep{caprioli14a} when the background magnetic field is parallel or~quasi-parallel.

\par
Likewise, a~fraction of the post-shock energy is converted into magnetic fields. If~$\epsilon_B$ represents this fraction, $\epsilon_B = u_{B}/u_{th}$, where $u_B = B^2/(8\pi)$ is the magnetic field energy density and~$B$ represents the magnetic field strength, respectively. Our knowledge about particle acceleration and magnetic field amplification is limited, as~both are highly complicated phenomena and require much higher computational facilities, which are currently not at hand (e.g., see \citep{caprioli14a,caprioli14b} for $B$ field amplifications in non-relativistic shocks explored using hybrid simulations). As~a result, $\epsilon_B$ is a poorly constrained~quantity. 

\par
Initially, the~SN ejecta remain in the free expansion phase until the reverse shock continues to exist in the outer power-law part of the ejecta. In~this phase, the~intensity of synchrotron radiation,  for a given frequency $\nu$, 
with synchrotron self-absorption (SSA) as the main absorption mechanism can be given~{by} 
\beq
I_{\nu}({\rm h}) = \frac{2 K_B T_b}{c^2 f(\frac{\nu_{peak}}{\nu_{abs}})} \frac{\nu^{5/2}}{\nu_{abs,0}^{1/2}} [1 - {\rm exp}(-\xi_{\rm h} \tau_{\nu_0})]
\label{eq_Inu}
\eeq
with~{\citep{kundu17}} 
\beq
f(x) = x^{0.5} [1- {\rm exp}(-x^{-{(p+4)/2}})]
\label{eq_f}
\eeq
for a shell of thickness $\Delta r$. Here, $\nu_{abs}$ is the absorption frequency for which the optical depth ($\tau_{\nu_{abs}}$) is one. $\xi_{\rm h} = \Delta s({\rm h})/2 \Delta r$, represents the path length, normalized by the thickness of the shell, along which the radiations travel along the line of sight. The~contribution of the 
emission from the entire shell is obtained by varying ${\rm h}$ between 0 and 1 ($0 \leq {\rm h} \leq 1$). For~${\rm h} = 0$, $\nu_{abs} = \nu_{abs,0}$, and $\tau_{\nu} = \tau_{\nu_0}$, with $\tau_{\nu}$ being the optical depth. Here, $K_B$ and $T_b$ are Boltzmann constant and brightness temperature, respectively. $\nu_{peak}$ is the radiation peak~frequency. 

\par 
The radio luminosity  of  
a shell with radius $r_s$ is $L_{\nu} = 8 \pi^2 r_s^2 \int^1_0 I_{\nu}({\rm h}) ~{\rm h} ~ d{\rm h}$, which can be written as {\citep{kundu_thesis_2019}}
\beq
L_{\nu} = 8 \pi^2 r_s^2 \vartheta_{\nu} ~ \frac{K_B T_b}{c^2 f(\frac{\nu_{peak}}{\nu_{abs}})} ~  \frac{\nu^{5/2}}{\nu_{abs,0}^{1/2}} ~ \big(1 - {\rm exp}(-\tau_{\nu_0}) \big)
\eeq
where $\vartheta_{\nu} = L_{\nu}/L_{\nu,0}$, with $L_{\nu,0} = 4 \pi^2 r_s^2 I_{\nu}(0)$. In~the optically thin regime, i.e.,~$\tau_{\nu} < 1$, the~luminosity is reduced to 
\beq
L_{\nu,\rm thin} = \frac{8 \pi^2 K_B T_b \vartheta_{\nu} r_s^2}{c^2 f(\frac{\nu_{peak}}{\nu_{abs}})} ~ \nu_{abs,0}^{(p+3)/2} ~ \nu^{-(p-1)/2}
\eeq
with {\citep{kundu17}}
\beq 
 \nu_{abs,0} = \bigg(2 \Delta r ~ \varkappa(p) ~ N_0 ~ B^{(p+2)/2} \bigg)^{(2/(p+4))}
\eeq
where $\varkappa(p)$ is the coefficient for the SSA. For~$p=3$ and a wind-like ambient medium, i.e.,~for $s = 2$, $L_{\nu,\rm thin}$ decreases with $t$ as
\beq
L_{\nu,\rm thin} \propto \varphi_{\nu} ~ \bigg (\frac{n-3}{n-2}\bigg)^{3.86} ~ (\dot M/v_w)^{\frac{1.93n - 8.43}{n-2}} ~ t^{-\frac{(n+2.57)}{n-2}}
\eeq
and for a constant-density medium ($s=0$), the~optically thin luminosity is {\citep{kundu17}}
\beq 
L_{\nu,\rm thin} \propto  \varphi_{\nu} \bigg (\frac{n-3}{n}\bigg)^{3.86} ~ (n_{ISM})^{\frac{1.93n - 8.43}{n}} ~ t^{\frac{(2.86n - 25.3)}{n}}
\eeq
 where $\varphi_{\nu} = T_b ~ \epsilon_{\rm rel}^{1.71} ~ \epsilon_B^{1.07} ~ \nu^{-1}$.

\par 
For stars with a 
higher mass loss prior to their explosions, the~density of the ambient medium is high. This medium is expected to be ionized due to the emission of strong ultraviolet radiation at the time of shock breakout and X-ray radiation from the shocked region \citep{fransson96}. As~a result, the~synchrotron radiation will suffer from external free--free absorption. If~we assume that the CSM is made up of hydrogen and helium with solar abundances, the~free--free absorption coefficient can be given as {\citep{rybicki79, kundu19}}
\beq
\alpha_{ff}(r) = 0.018 ~ T_{CSM}^{-1.5} ~ \frac{n_e^2(r)}{1+2y} ~ \nu^{-2} \bigg[\Bar{g}^H_{ff}(T_{CSM},\nu) +  4y ~ \Bar{g}^{He}_{ff}(T_{CSM},\nu)     \bigg]  {\rm cm^{-1}}
\eeq
where the~temperature of the CSM is represented by $T_{CSM}$. Here, $n_e(r)$ is the density of electrons in the ambient medium and $y$ is the ratio between the density of the helium atom to that of the hydrogen atom. $\Bar{g}^H_{ff}(T_{CSM},\nu)$ and $\Bar{g}^{He}_{ff}(T_{CSM},\nu)$ represent the velocity average Gaunt factor for hydrogen and helium, respectively. The~temperature of the ambient medium is found to be around $10^5$ K \citep{bjornsson14}. For~$T_{CSM} < 3 \times 10^5 Z^2$, where $Z$ represents the atomic number of a given element, and~radiation in the GHz frequency regime, which is the case for radio emission, $\Bar{g}_{ff}(T_{CSM},\nu) = \frac{\sqrt{3}}{\pi} ~ \bigg(17.7 + log{\bigg(\frac{T_{CSM}^{1.5}}{z\nu}\bigg)} \bigg)$. In~the case of $T_{CSM} > 3 \times 10^5 Z^2$, $\Bar{g}_{ff}(T_{CSM},\nu) = \frac{\sqrt{3}}{\pi} ~ \bigg(log{\bigg(\frac{K_B ~ T_{CSM}^{1.5}}{h\nu}\bigg)} \bigg)$ \citep{tucker75}, where $h$ is the Planck~constant.    

\par
When the external free--free absorption is important, the~radio luminosity becomes $L_{\nu}~{\rm exp}(-\tau_{ff} r_s)$, where $\tau_{ff}$ is the free--free optical depth. In~the case of wind-like ionized CSM that prevail up to a radius $r \gg r_s$, $\tau_{ff} = \frac{1}{3} r_s ~ \alpha_{ff}$.

\section{Observing SNe at Radio~Wavelengths}
\label{rad_sne}
Though all types of stars go through mass-loss episodes during their lifetimes, we are successful in detecting radio emissions from around 30\% of core-collapse SNe \citep{Bietenholz21a}.
In the case of Type Ia SNe (i.e., thermonuclear explosions), only very recently was a thermonuclear explosion, SN 2020eyj, detected after trying to detect these events at radio frequencies over decades of observation campaigns \citep{kool23}.  
In the following subsections, we discuss the properties of radio emission from different types of core-collapse and Type Ia~SNe.      

\subsection{Core-Collapse~SNe}
\label{rad_sne_cc}
\begin{figure}[H]
\includegraphics[width=0.65\textwidth]{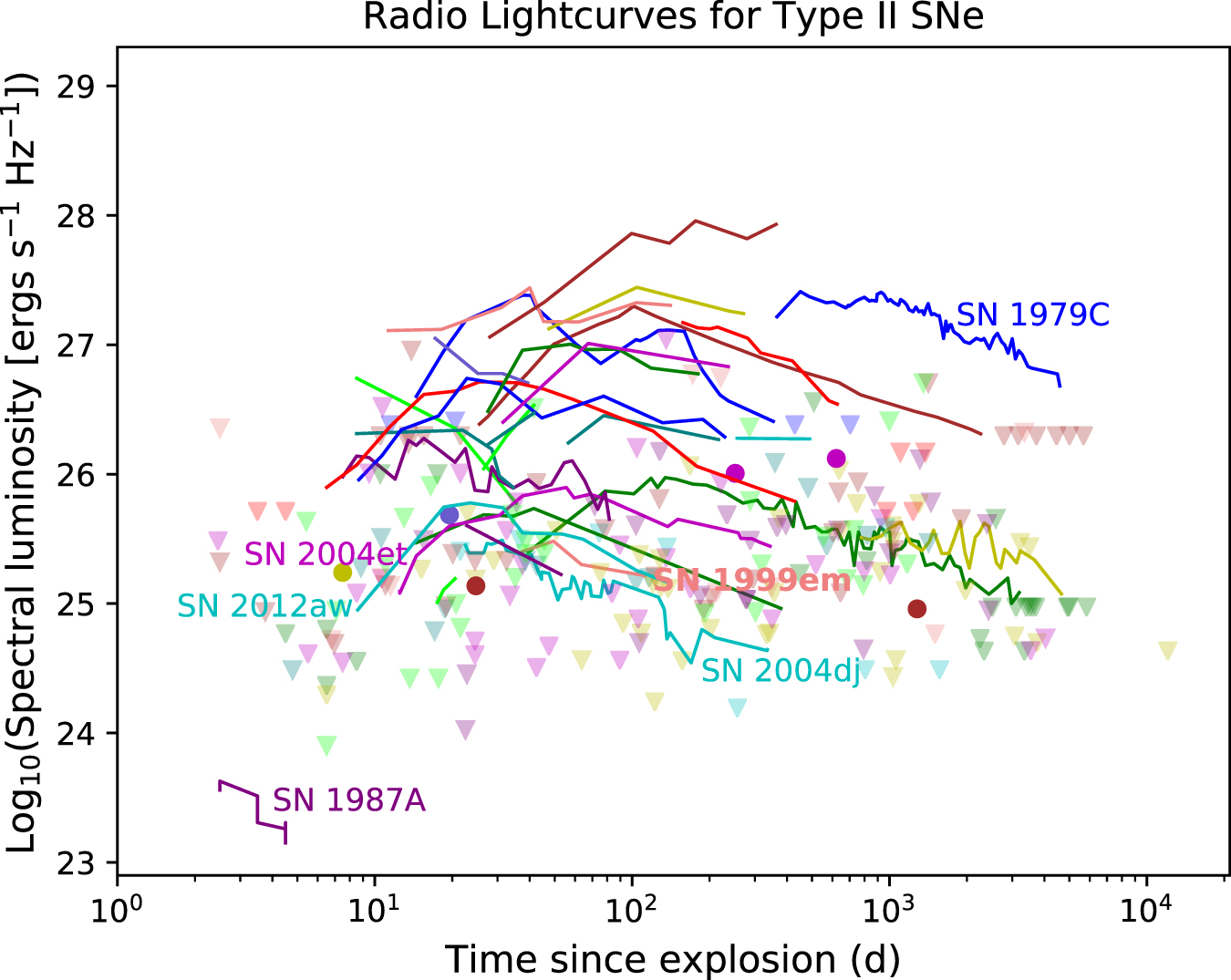}
\caption{{Evolution} 
 of luminosities of different types of Type II explosions at 4--10 GHz, at~a distance of less than 100 Mpc. The~light curve of Type IIP SN 1987A is at 2.3 GHz. The continuous lines that connect different measurements of a given explosion represent SNe with multiple detections. The solid circles show single detections, and the triangles represent the upper limits..Figure reproduced from \citep{Bietenholz21a} with~permission.}
\label{bitenholz_sn_ii_rad_light_curve}
\end{figure}

\begin{figure}[H]
\includegraphics[width=0.65\textwidth]{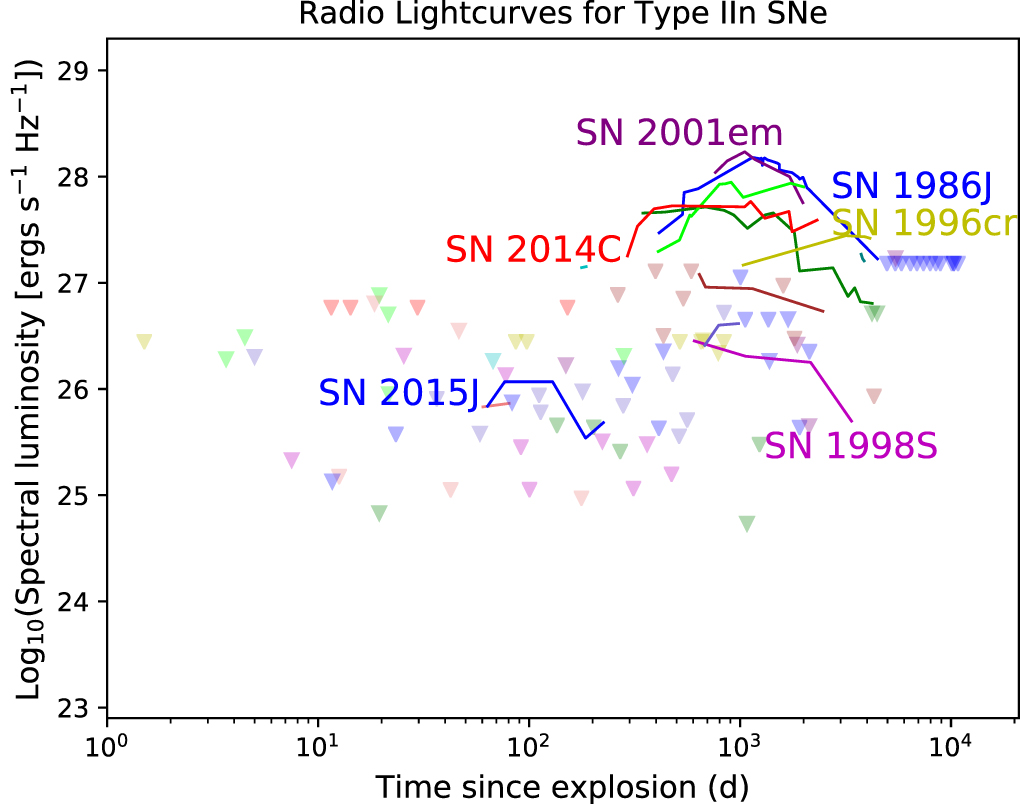}
\caption{Evolution 
 of luminosities of different types of Type IIn explosions at 4--10 GHz, at~a distance of less than 100 Mpc. See the caption of Figure \ref{bitenholz_sn_ii_rad_light_curve} for the details about lines and symbols used in this diagram. Figure reproduced from \citep{Bietenholz21a} with~permission.}
\label{bitenholz_sn_iin_rad_light_curve}
\end{figure}

\begin{figure}[H]
\includegraphics[width=0.45\textwidth]{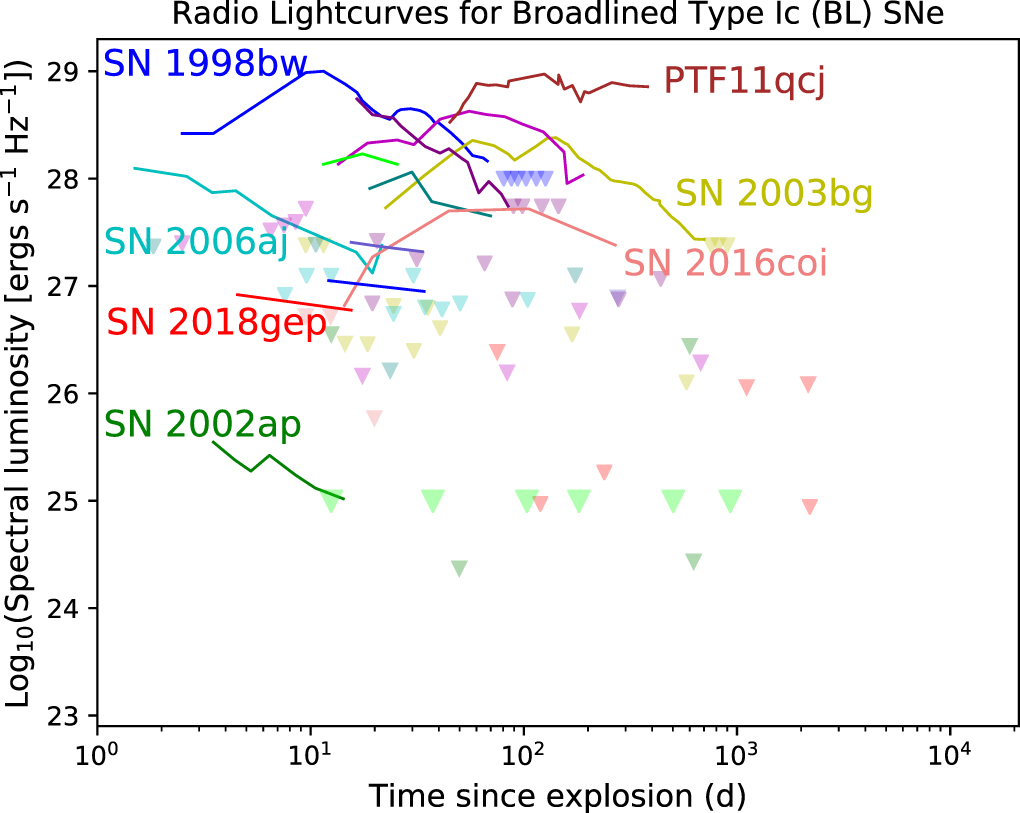}
\includegraphics[width=0.45\textwidth]{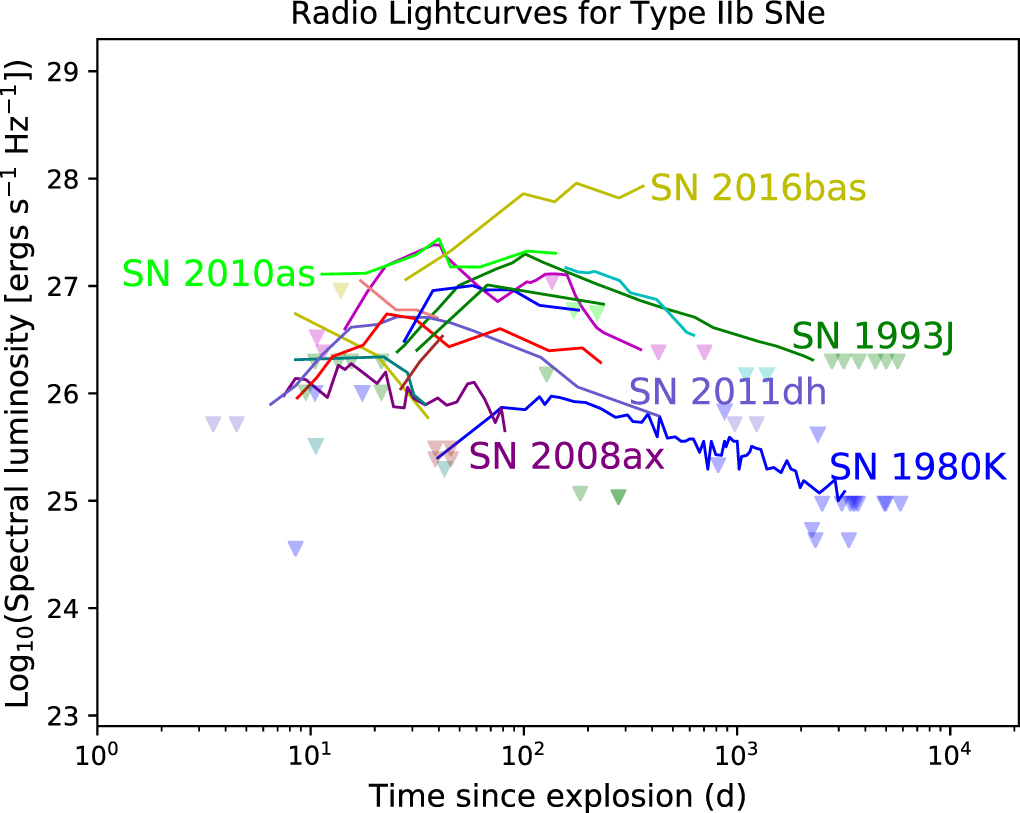}
\caption{{Light} 
 curves of different types of Type Ic (broad line) and Type IIb are displayed in the left and right panels at 4--10 GHz. For~Type Ic, explosions beyond 100 Mpc are included in this diagram, whereas for Type IIb, SNe within 100 Mpc are taken into consideration. For the details about lines and symbols used in this diagram, see the caption of Figure \ref{bitenholz_sn_ii_rad_light_curve}. Figure reproduced from \citep{Bietenholz21a} with~permission.}
\label{bitenholz_sn_ic_iib_rad_light_curve}
\end{figure}

\begin{figure}[H]
\includegraphics[width=0.9\textwidth]{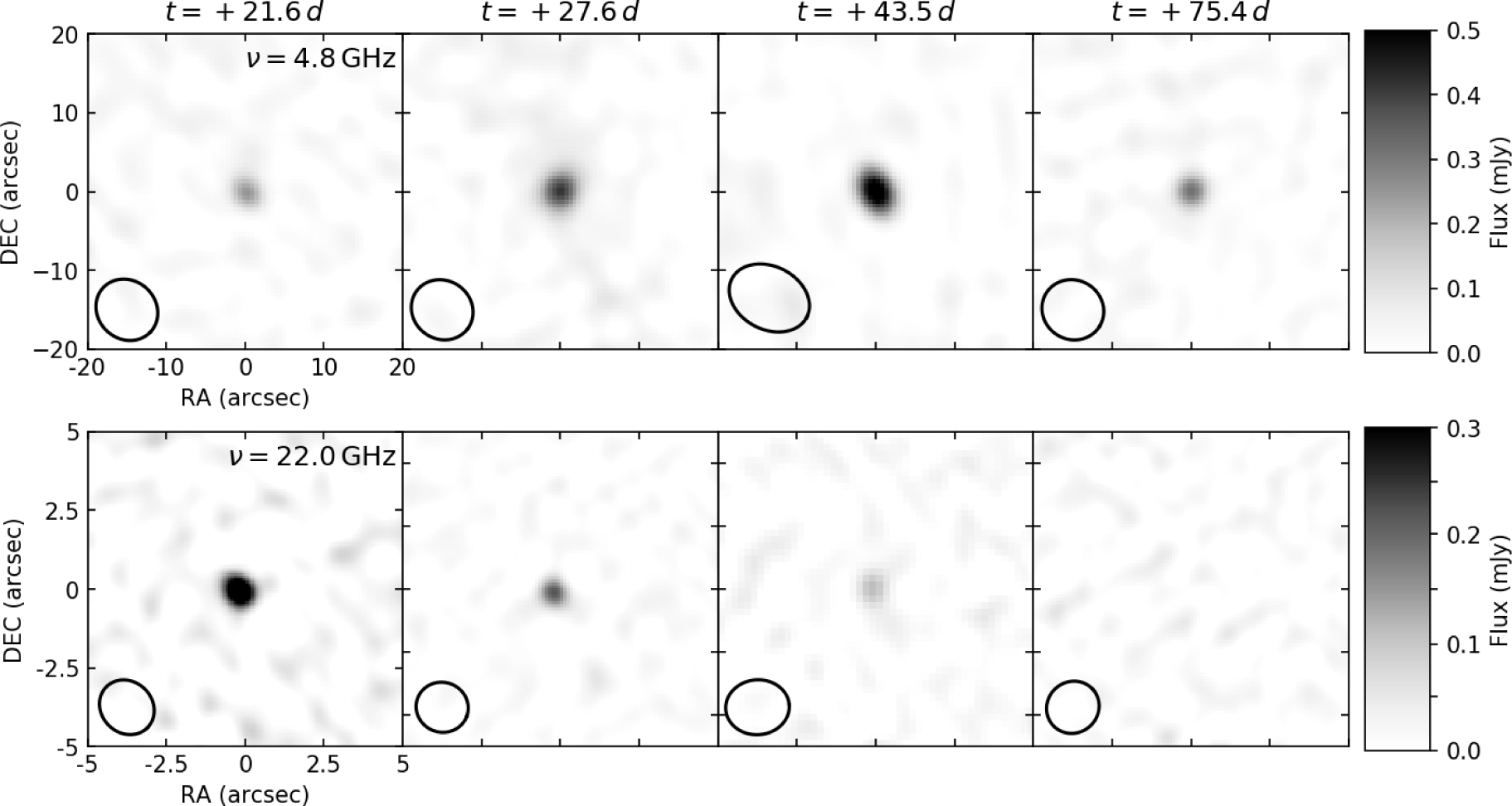}
\caption{{VLA} 
 observations of SN 2016X from 21 days to 75 days at 4.8 GHz and 22 GHz. Figure reproduced from \citep{Ruiz-Carmona22} with permission. For~each image, the~secondary beam size is different, as~shown on the bottom left of each of them.
 }
\label{SN2016X_radio_VLA_image}
\end{figure}

\begin{figure}[H]
\includegraphics[width=0.65\textwidth]{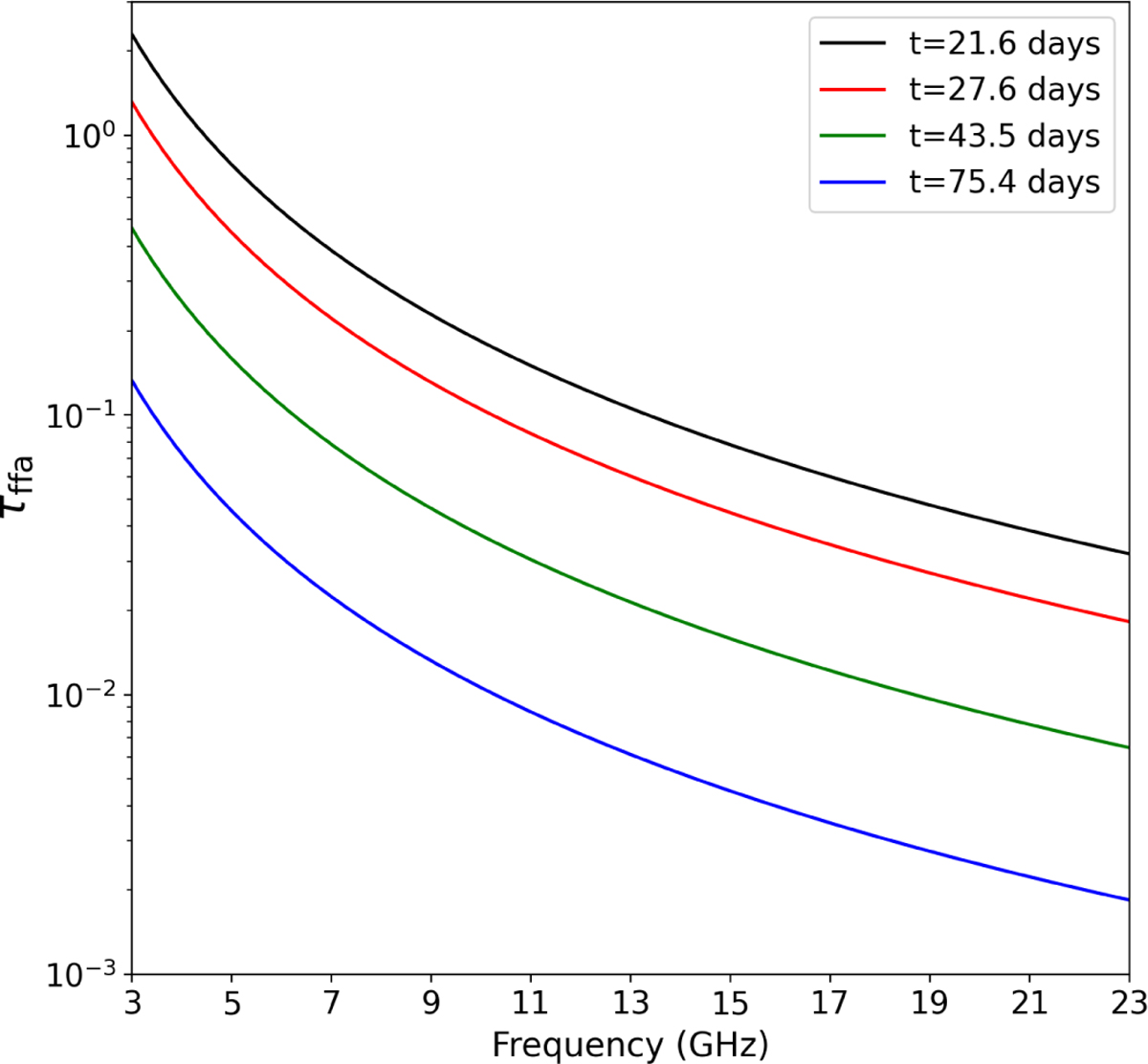}
\caption{ Evolution of optical depth  with frequency at four different epochs for SN 2016X. Figure reproduced from \citep{Ruiz-Carmona22} with permission.
}
\label{SN2016X_opt_dpt_evlv}
\end{figure}

As mentioned, in~the case of core-collapse SNe, the~ambient media often possess a wind-like density profile, which decreases with radius. However, there are significant variations in their light curves, as the mass loss is uncertain and not very well understood before the explosion of the stars. The~light curves of different types of core-collapse SNe, at~4--10 GHz frequencies, are shown in {Figures~\ref{bitenholz_sn_ii_rad_light_curve}, \ref{bitenholz_sn_iin_rad_light_curve}, and \ref{bitenholz_sn_ic_iib_rad_light_curve}}
. Here, Figure~\ref{bitenholz_sn_ii_rad_light_curve} demonstrates the evolution of spectral luminosity for 106 Type II SNe (excluding Type IIn), that are located at less than 100 Mpc. The~light curve of Type IIP explosion SN 1987A is shown here at 2.3 GHz. Among~all core-collapse explosions, a~significant percentage of these were found to belong to the Type IIP category \citep{Smartt09}. However, the~similarities between their progenitors are less pronounced. During~the plateau phase of Type IIP event SN 2016X, the~VLA radio observations at 4.8 and 22 GHz revealed that in the self-similar scenario, to~explain the detections, a~significantly higher amount (almost 30 times more) of the shock energy is required to be channeled to electrons compared with that in magnetic fields \citep{Ruiz-Carmona22}. The~VLA observations of SN 2016X from 21 days to 75 days at 4.8 GHz and 22 GHz are shown in Figure~\ref{SN2016X_radio_VLA_image}. As~expected, the~SN becomes brighter at the lower frequency at a later phase, around 44 days, whereas at 22 GHz, it fades away and~was not detected at the last epoch of observation. The~evolution of the free--free optical depth as a function of frequency for four different epochs is displayed in Figure~\ref{SN2016X_opt_dpt_evlv}, which decreases with frequency over~time.

\unskip

\unskip

\unskip

\unskip

\begin{figure}[H]
\includegraphics[width=13.7cm,angle=0]{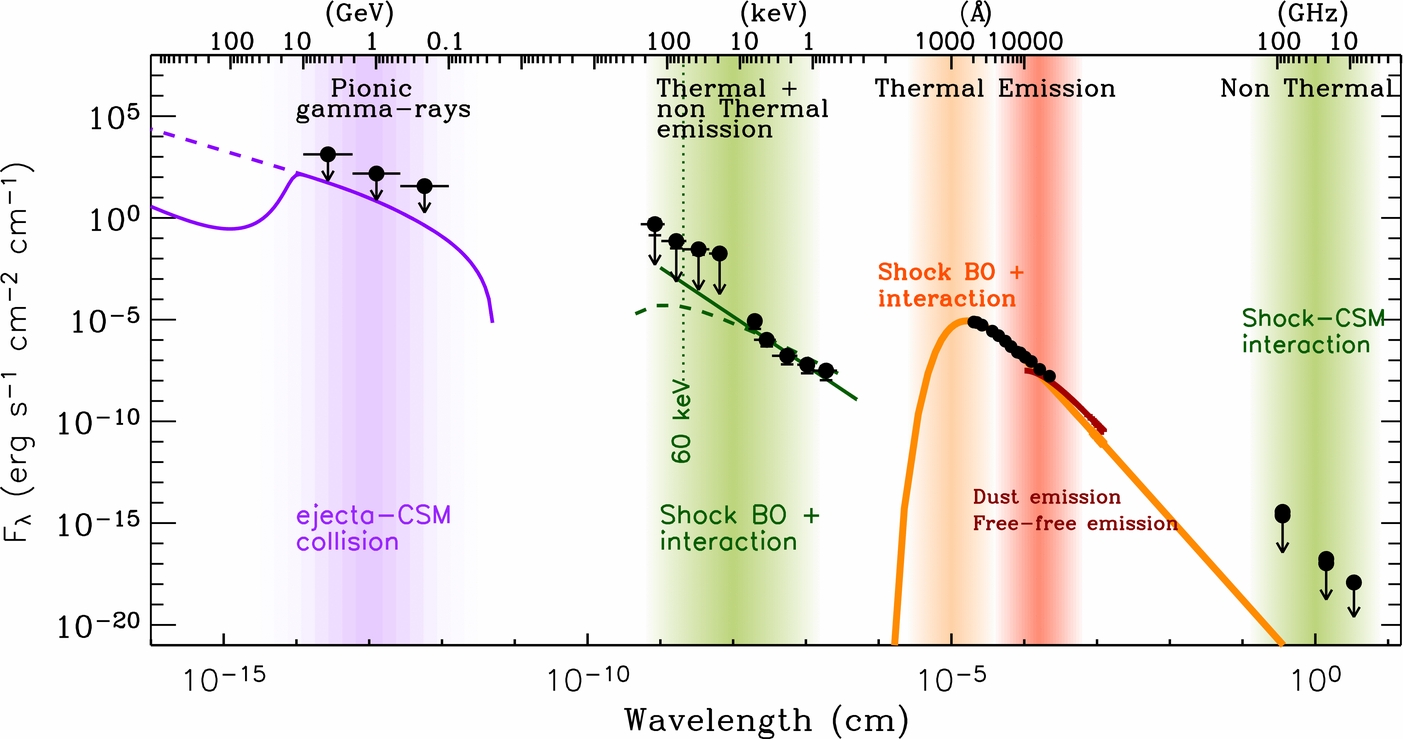}
\caption{{The} 
 SED of SN 2009ip around the optical peak, 30 September 2012, from radio to gamma rays. At~radio wavelength, the~SN is not detected as a result of free--free absorption by the external high-density material. Figure reproduced from \citep{Margutti14} with permission. 
}
\label{SN_2009ip_SED}
\end{figure}

\begin{figure}[H]
\includegraphics[width=10cm,angle=0]{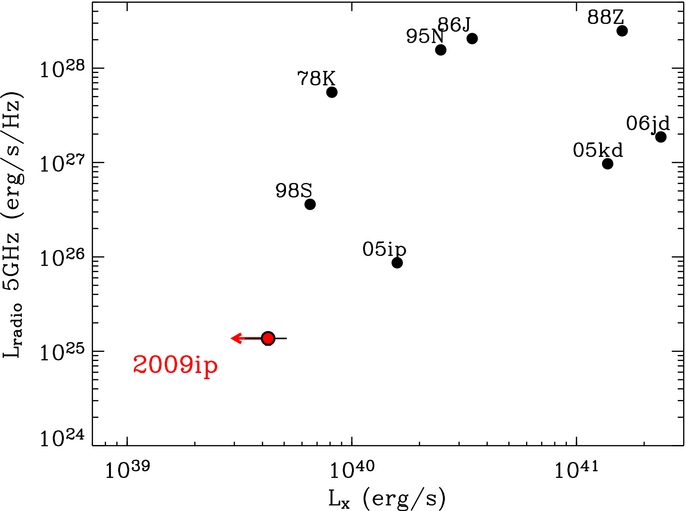}
\caption{
{Radio} 
 luminosity at 5 GHz peak and X-ray luminosity at radio peak for SN 2009ip along with X-ray-bright Type IIn SNe. As~demonstrated here, SN 2009ip (represented by a red dot) is not a strong radio and X-ray emitter, like other IIn events, shown as black dots. At its radio peak, SN 2009ip was not detected at X-ray frequencies, as indicated by the red arrow.  
Figure reproduced from~\citep{Margutti14} with permission.
}
\label{SN_IIn_radio_Xray_corr}
\end{figure}

Among core-collapse SNe, the~Type IIn progenitors eject a substantial amount of matter prior to their explosions. The~light curves of a handful of Type IIn SNe are displayed in Figure~\ref{bitenholz_sn_iin_rad_light_curve}. In~most cases, these SNe start to be seen at radio wavelengths at a much later epoch, and~they often remain visible for a longer duration compared with other types of core-collapse explosions. A~peculiar double-exploded Type IIn event is SN 2009ip. This explosion raised significant doubts about the explosion of massive stars and their mass-loss episodes \citep{Mauerhan13,Prieto13,Margutti14,Smith22}. The~spectral energy distribution (SED) of this explosion around the optical peak (30 September 2012) during the double-explosion episode in 2012 is shown in Figure~\ref{SN_2009ip_SED} from radio to gamma rays. At~radio wavelengths, the~SN is not detected as a result of free--free absorption by the external high-density material. Compared with other X-ray-bright IIn events, SN 2009ip is a weak emitter of X-rays \citep{Chevalier87,van Dyk93,Fabian96,Houck98,Pooley02,Zampieri05,Smith07,Pooley07,Chandra07,Stritzinger07,Chandra09,Chandra12,Margutti14}, as~shown in Figure~\ref{SN_IIn_radio_Xray_corr}.

\unskip

\begin{figure}[H]
\includegraphics[width=0.8\textwidth]{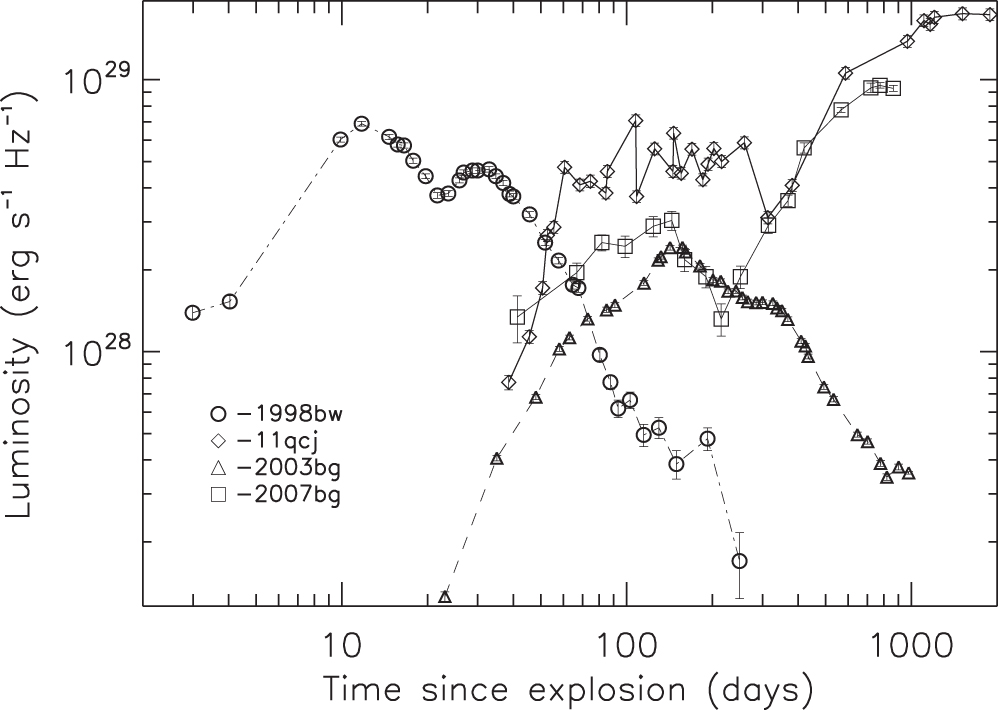}
\caption{{Spectral} 
  luminosity variation with time at 5 GHz for two Type Ic-BL explosions, SNe 1998bw and PTF11qcj, along with that of Type Ibc/IIb SNe, SN 2003bg and 2007bg. Figure reproduced from~\citep{Palliyaguru19} with permission.
}
\label{IcBL_SN_PTF11qcj}
\end{figure}

\begin{figure}[H]
\includegraphics[width=0.75\textwidth]{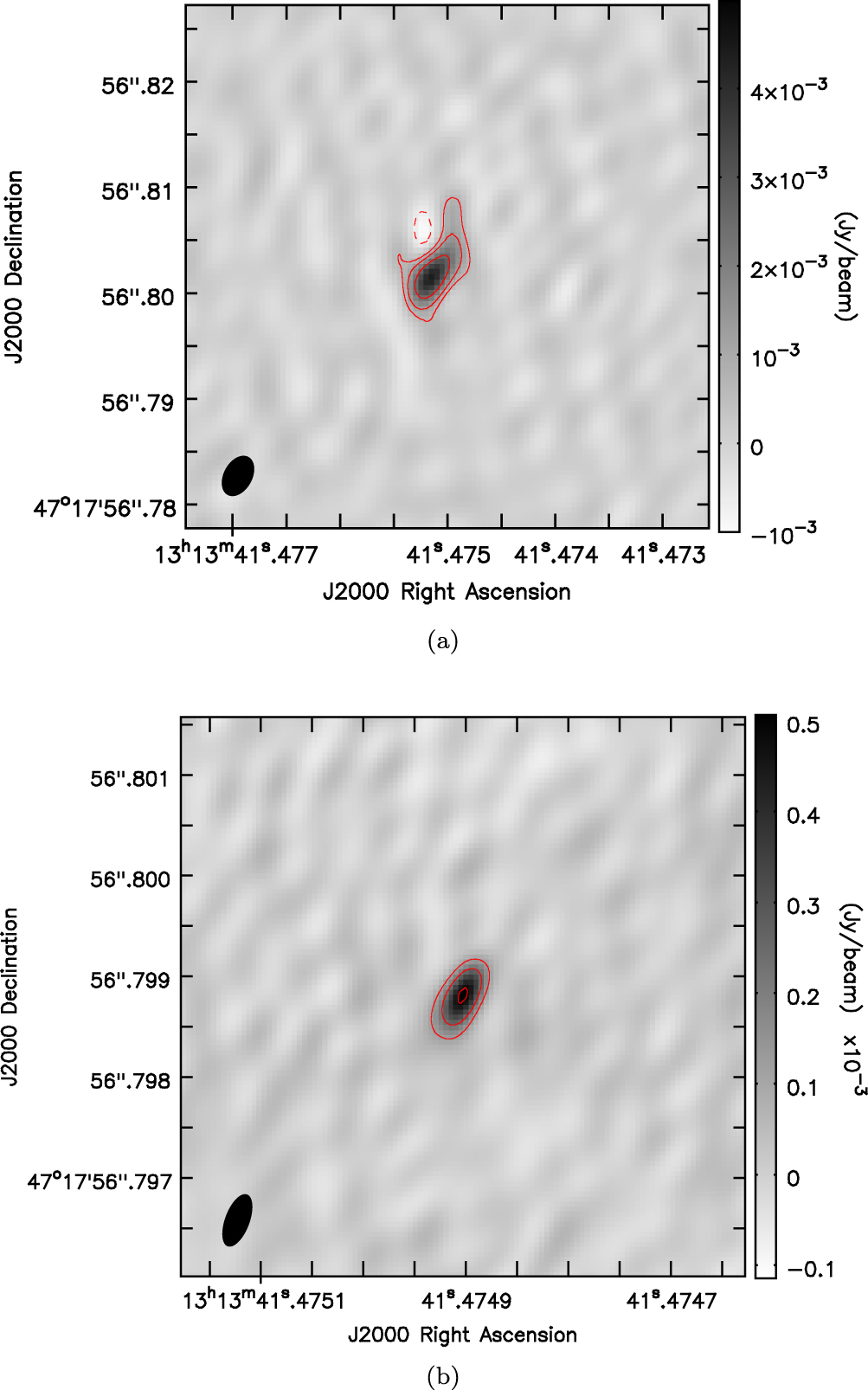}
\caption{{The} 
 VLBI observations of Type Ic-BL PTF11qcj at (\textbf{a}) 1.66 GHz and (\textbf{b}) 15.37 GHz~around 7.5 years after the explosion. The~rms noises at these frequencies are $125$ {$\mu$Jy/beam} 
 and $24$ {$\mu$ Jy/beam}, respectively. The~contours here represent ($-5, 5, 10, 20$) times the rms noise at the corresponding frequencies. 
Figure reproduced from \citep{Palliyaguru21} with permission.
}
\label{VLBI_IcBL_SN_PTF11qcj}
\end{figure}

\begin{figure}[H]
\includegraphics[width=0.7\textwidth]{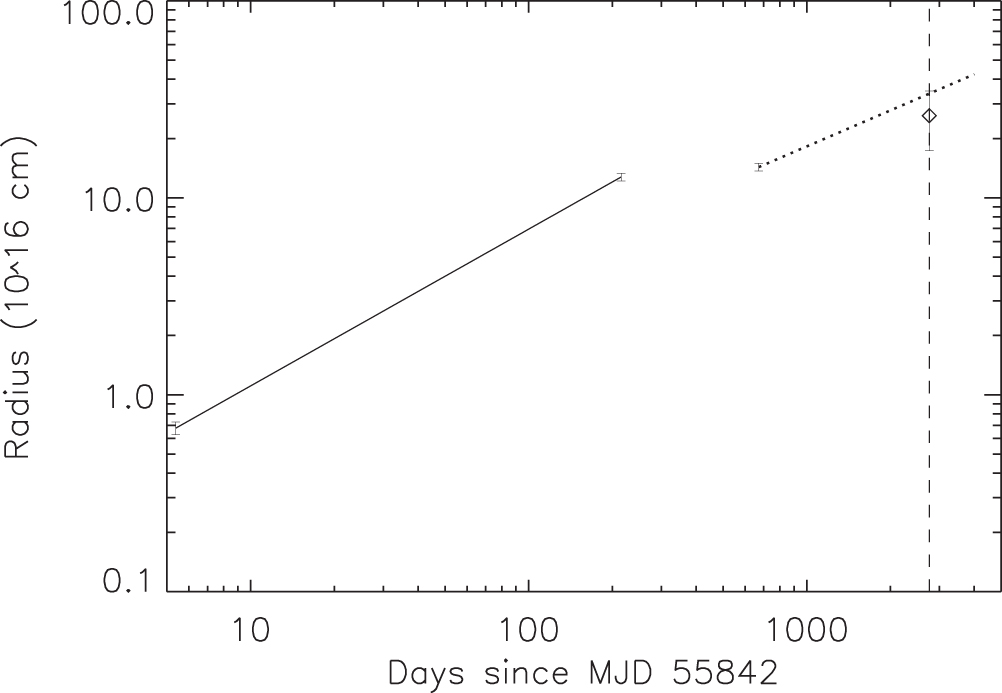}
\caption{The radial expansion of SN PTF11qcj as~a function of time. The~dashed line represents the 16 GHz VLBI observations.
The VLBI 15 GHz detection, displayed by a diamond, predicts a radius of around $3 \times 10^{17}$ cm, which implies an angular diameter of around 0.4 mas. The~shock is decelerated during its rebrightening phase. Figure reproduced from \citep{Palliyaguru21} with permission. 
}
\label{SN_PTF11qcj_radius_expansion}
\end{figure}

In Figure~\ref{bitenholz_sn_ic_iib_rad_light_curve}, the~light curves of the fully (Type Ic) or partially (Type IIb) stripped-envelope SNe are shown in the frequency window of 4--10 GHz. While for Type IIb, only SNe within 100 Mpc are included, for~Type Ic (broad line; BL), those beyond this distance are also considered. As~illustrated in the left panel of Figure~\ref{bitenholz_sn_ic_iib_rad_light_curve}, SNe 1998bw and PTF11qcj have the highest spectral luminosities among all explosions observed to date. Additionally, the~latter explosion continues to be brighter for a significantly longer duration. The~early light curves of the Type Ic-BL explosions, SNe 1998bw and PTF11qcj, have revealed two radio peaks, as~demonstrated in Figure~\ref{IcBL_SN_PTF11qcj}. However, the~second peak for PTF11qcj is brighter than the first peak, which is in contrast to what was observed for SN 1998bw, which is linked to a long Gamma-Ray Burst (GRB) event. While the second radio-bright peak could be due to a variation in the density of the CSM, it may indicate the presence of a GRB off-axis jet \citep{Palliyaguru19}. To~further investigate the possibility of the association of the radio rebrightening of PTF11qcj with a GRB jet, this event was observed with Very-Long-Baseline Interferometry (VLBI) at 1.66 GHz and 15.37 GHz around 7.5 years after the explosion, which is shown in Figure~\ref{VLBI_IcBL_SN_PTF11qcj}. At~the submillisecond level, the~unresolved radio-emitting ejecta suggest the absence of relativistically ejected material \citep{Palliyaguru21}, implying that this explosion is unlikely to be linked to an off-axis GRB event~\citep{Palliyaguru21}. As~a result, the~enhancement in radio flux could be due to an interaction with a high-density ambient medium. The~VLBI observations of PTF11qcj predict a radial expansion of around $3 \times 10^{17}$~cm, which implies an angular diameter of around 0.4 mas \citep{Palliyaguru21}. The~radial expansion of this SN with time, as~shown in Figure~\ref{SN_PTF11qcj_radius_expansion}, revealed a decelerated shock during the time of its second radio peak. This feature is found to be similar to that of SN 2014C, which was originally classified as a Type Ib explosion, but~at a later stage, it emerged as a Type IIn event when it started to interact with a dense CSM ejected by the progenitor before the explosion \citep{Milisavljevic15,margutti17,Bietenholz21b}. 

\unskip

\begin{figure}[H]
\includegraphics[width=0.75\textwidth]{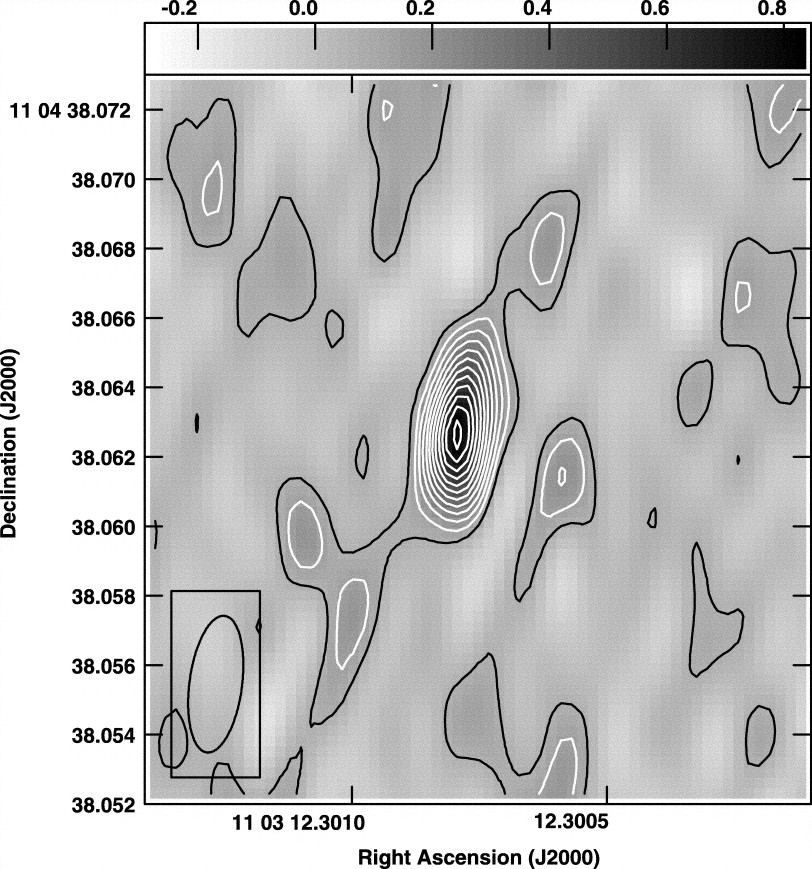}
\caption{{The} 
 VLBA observations of Type Ibc SN 2003L at 4.9 GHz around 65 days after the explosion. The~flux density at this frequency is found to be $848 ~\pm 64$ $\mu$Jy. This detection implies that the SN did not expand beyond a radius of $9.1 \times 10^{17}$ cm, suggesting the presence of subrelativistic ejecta for this event. The~contours here represent an increment of 1$\sigma$ in flux density which is around 67 $\mu$Jy.  
Figure reproduced from \citep{soderberg2005} with permission. 
}
\label{VLBI_SN2003L_Ic_GRB}
\end{figure}

\begin{figure}[H]
\includegraphics[width=0.6\textwidth]{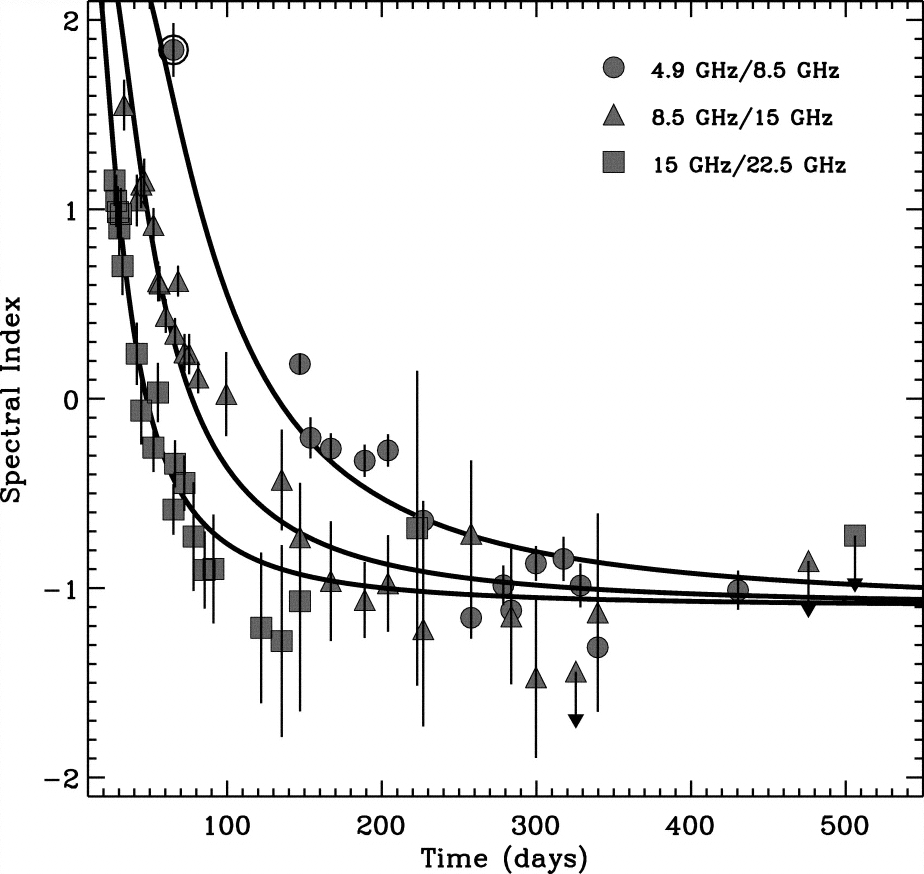}
\caption{{Evolution} 
 of spectral indices of SN 2003L as a function of time at 4.9 GHz/8.5 GHz, 8.5~GHz/15 GHz, and 15 GHz/22.5 GHz displayed as solid circles, triangles, and~squares, respectively. The~circled dot represents the VLBA observation. The~solid lines are the model predictions of the SSA model discussed in \citep{soderberg2005}. Figure reproduced from \citep{soderberg2005} with permission.   
}
\label{SN2003L_spectral_index_evoltn}
\end{figure}

Another stripped-envelope supernova that has radio luminosity comparable to that of SN 1998w is Type Ibc SN 2003L \citep{soderberg2005}. The~Very-Long-Baseline Array (VLBA) observations of this SN at 4.9 GHz around 65 days after the explosion are shown in 
Figure~\ref{VLBI_SN2003L_Ic_GRB}. At~this frequency, this event was detected with a flux density of $848 ~ \pm ~ 64 $ {$\mu$Jy}. This detection implies that around 65 days  after the explosion, the~SN did not expand beyond a radius of $9 \times 10^{17}$ cm, suggesting the presence of subrelativistic ejecta for SN 2003L, which is in contrast to SN 1998bw, for~which the radio-emitting shell is found to expand at mildly relativistic speed \citep{soderberg2005}. The~evolution of the spectral indices of this SN as a function of time is shown in Figure~\ref{SN2003L_spectral_index_evoltn}. In~the optically thin regime, the~spectral index is estimated to be around $-$1.1 \citep{soderberg2005}.

\unskip




\unskip

\begin{figure}[H]
\includegraphics[width=0.6\textwidth]{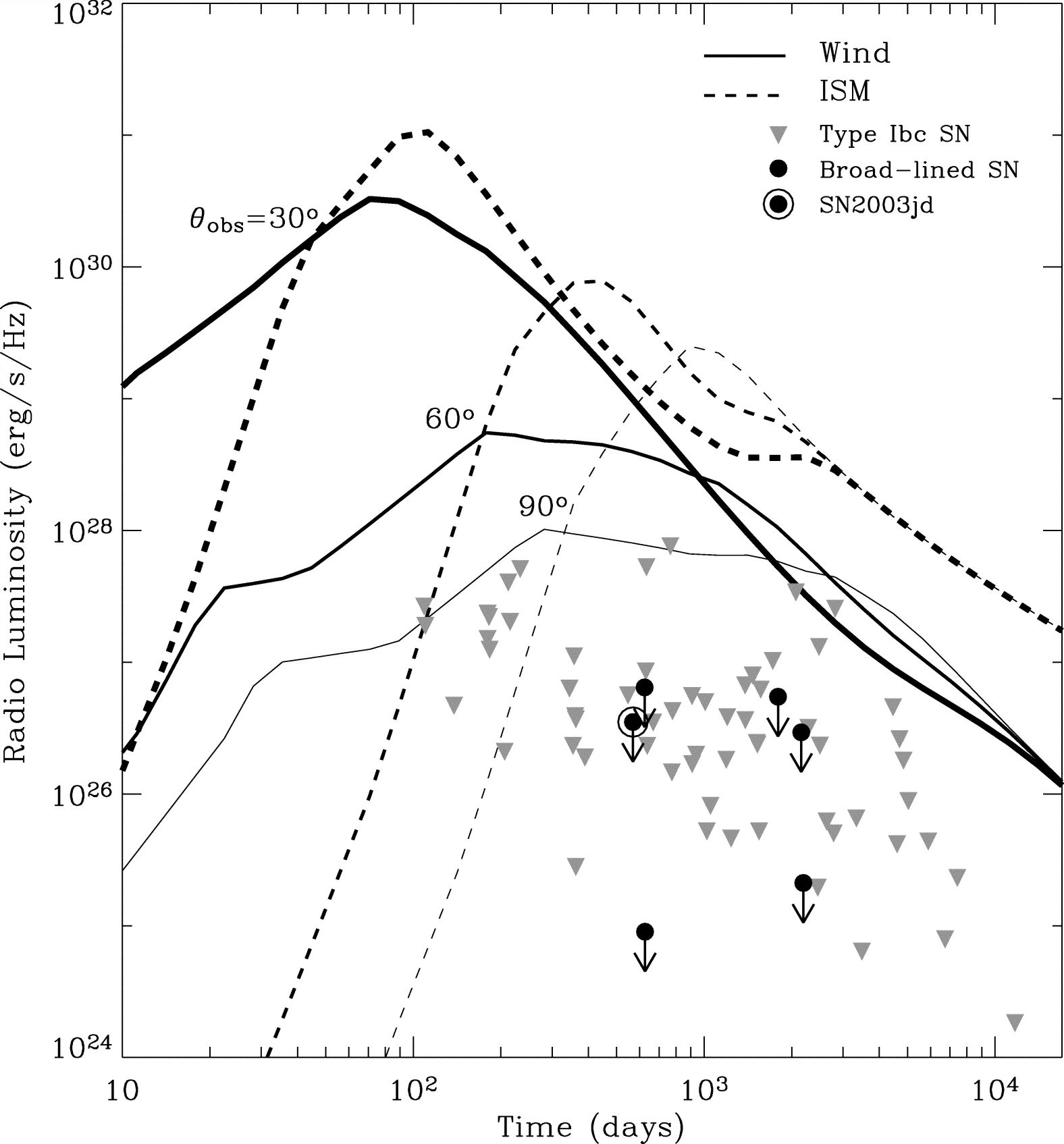}
\caption{{Late-time} 
 radio observations (3$\sigma$ upper limits) of 62 Ibc events, in filled faded triangles, and 6 broad-lined SNe, in filled circles with arrows, at 8.46 GHz, along with the  GRB afterglow models for both wind-like and ISM media for the viewing angles of 30 $~\!\!^\circ$, 60 $~\!\!^\circ$, and 90 $~\!\!^\circ$ from the initially collimated direction of the GRB jet. For~all these off-axis viewing angles, the models predict much higher radio luminosity for standard GRB parameters (where the~beam-corrected ejecta energy is $10^{51}$ erg; the~GRB jet opening angle $\theta_j = 5 ~\!\!^\circ$; $\epsilon_e = \epsilon_{\rm B} = 0.1$; for~a wind-like medium $v_w = 10^{3}$ \kms, $\mdot = 10^{-5}$ $\msunyr$; and~for an ISM medium, the particle density is assumed to be 1 cm$^{-3}$). Figure reproduced from \citep{Soderberg06} with permission.
}
\label{Ibc_SNe_68_GRB}
\end{figure}

Generally, stripped-envelope SNe are believed to be explosions in binary systems, and~it is the binary interactions that lead to the ejection of their complete or partial envelope~\citep{smith11}. Studies have shown that in~isolation, stars with zero-age main-sequence (ZAMS) masses of more than 20 $\msun$ can undergo significant mass loss due to strong winds~\citep{schaller11, vink05}. However, the~number of single massive stars is not high enough to match the rate of stripped-envelope SNe. In~addition, stars that end their lives with very low envelope masses can occur only for those within very shallow and specific ranges of their ZAMS masses \citep{smartt15}. A~small fraction of these stripped-envelope explosions are expected to be associated with long GRB events such as SN 1998bw \citep{MacFadyen01,Podsiadlowski04}. With~late-time radio observations of around 70 Type Ibc SNe, Soderberg~et~al. 2006c \citep{Soderberg06} established that at most, a few percent of Type Ibc events are likely to have been associated with off-axis GRB jets. Figure~\ref{Ibc_SNe_68_GRB} demonstrates late-time radio observations ((3$\sigma$ upper limits)) of these 68 Ibc events at 8.46 GHz, together with the GRB afterglow models for both wind-like and ISM media for the viewing angles of 30 $~\!\!^\circ$, 60 $~\!\!^\circ$, and 90 $~\!\!^\circ$ from the initially collimated direction of the GRB jet
. 
For all these off-axis viewing angles, the models predict much higher radio luminosity for standard GRB parameters (where the beam-corrected ejecta energy is $10^{51}$ erg; the~GRB jet opening angle $\theta_j = 5 ~\!\!^\circ$; $\epsilon_e = \epsilon_{\rm B} = 0.1$; for~a wind like medium $v_w = 10^{3}$ \kms, $\mdot = 10^{-5}$ $\msunyr$; and~for an ISM medium, the~particle density is assumed to be 1 cm$^{-3}$
). Although~the Very Large Array Intensive Study of Naked Supernovae (\citep{Berger02, Berger03}) program has observed a good number of Type Ibc SNe at radio wavelengths \citep{soderberg2005,Soderberg06a,Soderberg06b,Soderberg06,Soderberg08,Soderberg10a,Soderberg10b,wellon12}, to~establish a clear connection between GRB and stripped-envelope SNe, a~larger sample of hydrogen-poor supernovae should be studied at both early and late epochs at wideband radio wavelengths.

\begin{figure}[H]
\includegraphics[width=0.7\textwidth]{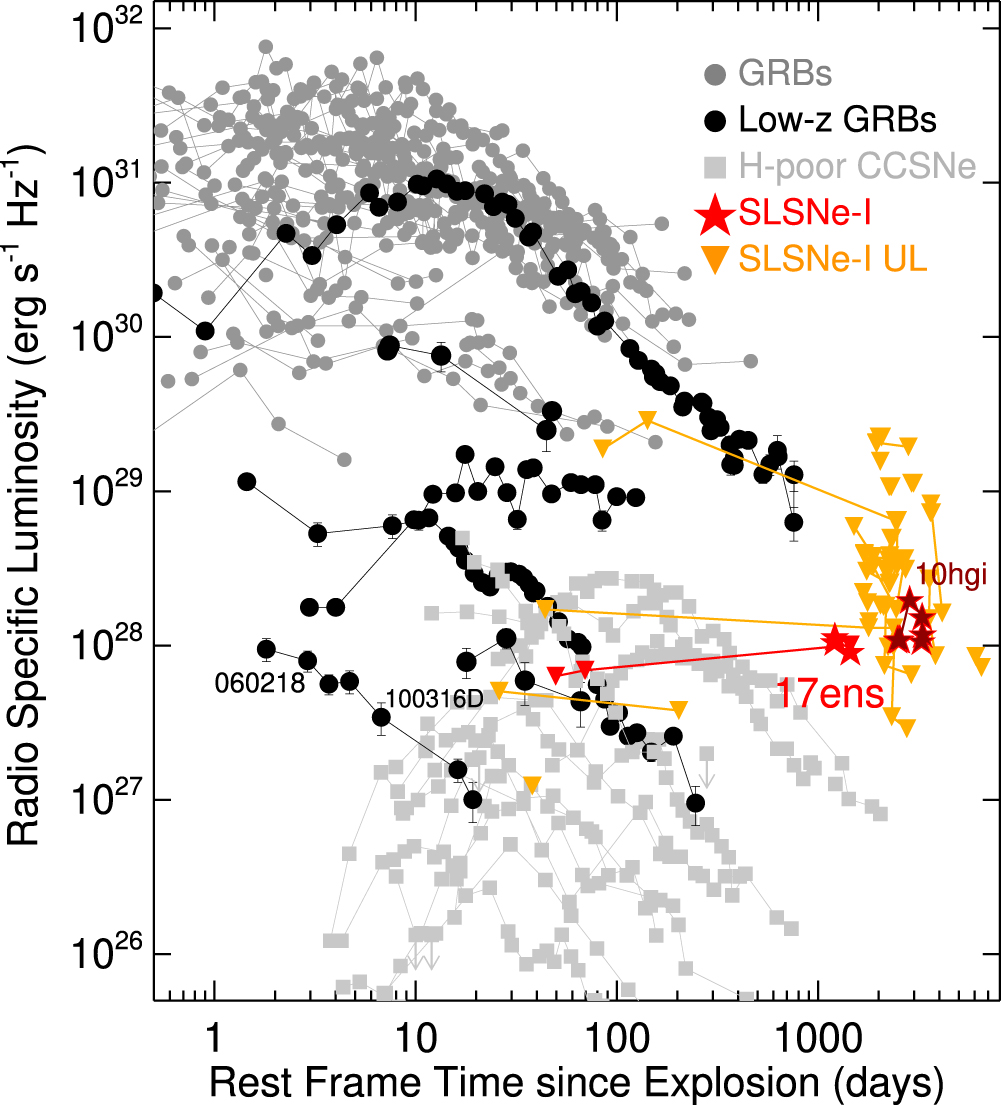}
\caption{{The} 
 radio light curves of Type I SLSNe and GRBs. The~triangles displays the radio upper limits for Type I SLSNe, while the red stars represents the radio detection that was obtained from SN 2017ens around 3.3 years after the explosion. Figure reproduced from \citep{Margutti23} with permission.
}
\label{SLSNe_GRB}
\end{figure}

Additionally,~study on a limited sample of early and late-time radio observations of hydrogen-poor superluminous (SL) SNe, which are at least ten times more luminous than normal core-collapse SNe \citep{chomuik11,Quimby11,gal-yam12,Moriya18}, reveals the absence of relativistic jets in the majority of Type I SLSNe. For~comparison, the~radio light curves of Type I SLSNe and GRBs are shown in Figure~\ref{SLSNe_GRB}. After~decades of effort, the~earliest radio detection yield from SLSN 2017ens was around 3.3 years after the explosion \citep{Margutti23}. The~early detection of radio emissions from such events is crucial to probing the presence of mildly relativistic jets in the early phase of the explosion, as~explained earlier.   

\unskip



\par
\begin{figure}[H]
\includegraphics[width=10cm,angle=0]{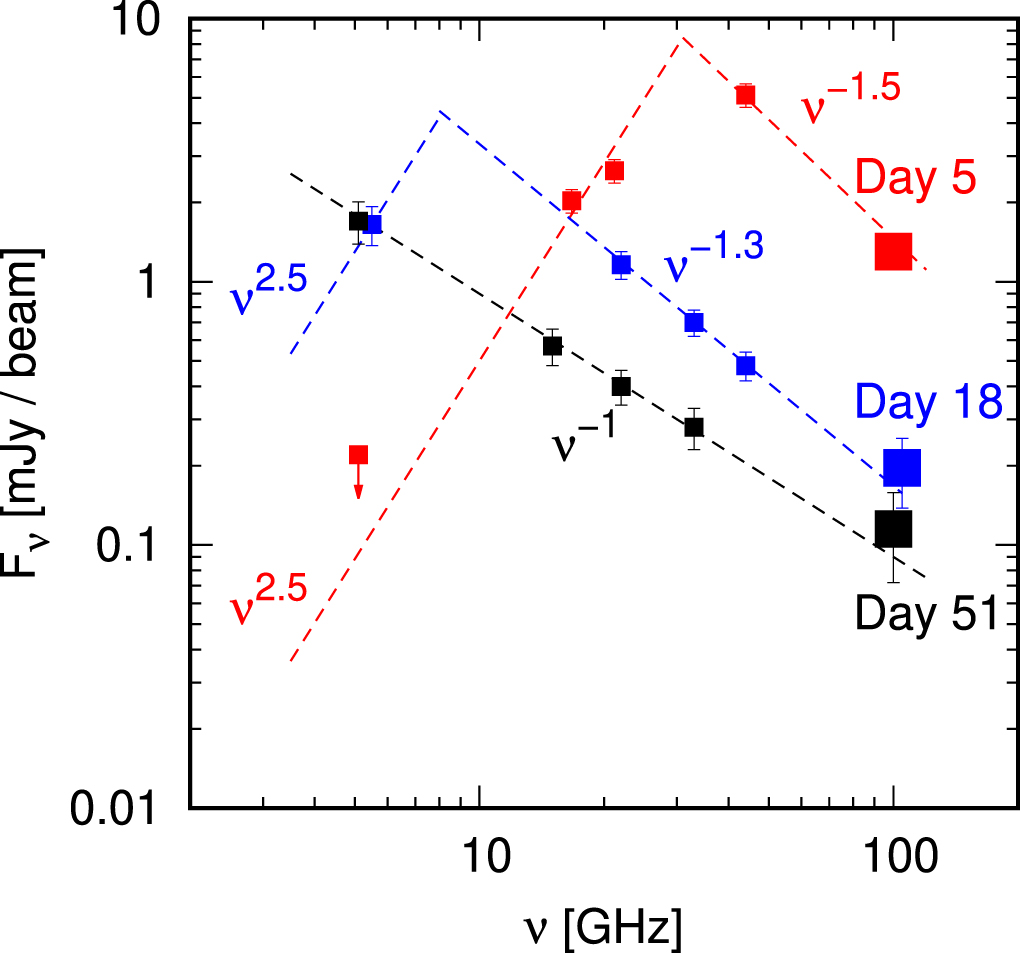}
\caption{{SEDs} 
 of SN 2020oi on day 5, day 18, and day 51. In~the optically thick regime, the~flux varies with frequency with a spectral index of 2.5, and~in the optically thin regime, it is in the range of $-1$ to $-2.5$. As~observed in the SEDs, on~day 5, the~SN just becomes optically thin at 44 GHz, whereas it is in the fully thin regime at 100 GHz by this time. Figure reproduced from \citep{keichii21} with~permission.}
\label{radio_sn2020oi_images}
\end{figure}

 
Furthermore, to~explore these hydrogen-poor SNe in the hundreds-of-GHz regime, the~mass-loss history of the pre-SN star of the nearby Type Ic explosion SN 2020oi was investigated by Keichii~et~al. in 2021 \citep{keichii21} using radio observations of the Atacama Large Millimeter/submillimeter Array (ALMA). The~results suggested a variable mass loss of the progenitor prior to the explosion with a time scale for each mass loss phase lasted less than a year. The~radio spectral energy distributions (SEDs) of the SN on day 5, day 18, and~day 51 after~the explosion are shown in Figure~\ref{radio_sn2020oi_images} as red, blue, and~black dashed lines, respectively. In~the optically thick regime, the~flux varies with frequency with a spectral index of 2.5, and~in the optically thin regime, it is in the range of $-1$ to $-2.5$. As~observed in the SEDs, on~day 5, the~SN just becomes optically thin at 44 GHz, whereas it is in the fully thin regime at 100 GHz by this time, which provides a stringent limit on the density of the CSM around the~SN.

The radio light curves of partially stripped hydrogen envelope explosions, called SNe of Type IIb, are shown in the right panel of Figure~\ref{bitenholz_sn_ic_iib_rad_light_curve}. Some Type IIb SNe have been found to remain bright at radio wavelengths for decades; one such SN is Type IIb SN 1993J, which was detected from a couple of weeks after the explosion up to several decades~\citep{bartel00,bartel02,bietenholz03, chandra04,weiler07,martividal11}. 
The almost spherical radio shell of this SN is well resolved by Very-Large-Baseline Interferometry (VLBI) (for details, see \citep{bartel02, bietenholz03}). Analyzing the pre-explosion images, Chen~et~al. 1995 \citep{cohen95} identified the progenitor star of SN 1993J. Another Type IIb SN, which has shown radio evolution similar to that of SN 1993J, is SN 2011dh (see the SN 2011dh light curve in the right panel of Figure~\ref{bitenholz_sn_ic_iib_rad_light_curve}).  The~other few SNe for which pre-SN stars are detected are SN 2008ax, SN 2011dh, and~SN 2013df \citep{crockett08, vandyk11, vandyk14}. In~addition, using ultraviolet emissions, the~binary companion of Type IIb SN 2001ig was recently detected by Ryder~et~al. in 2018 \citep{ryder18}.

\begin{figure}[H]
\includegraphics[width=0.78\textwidth]{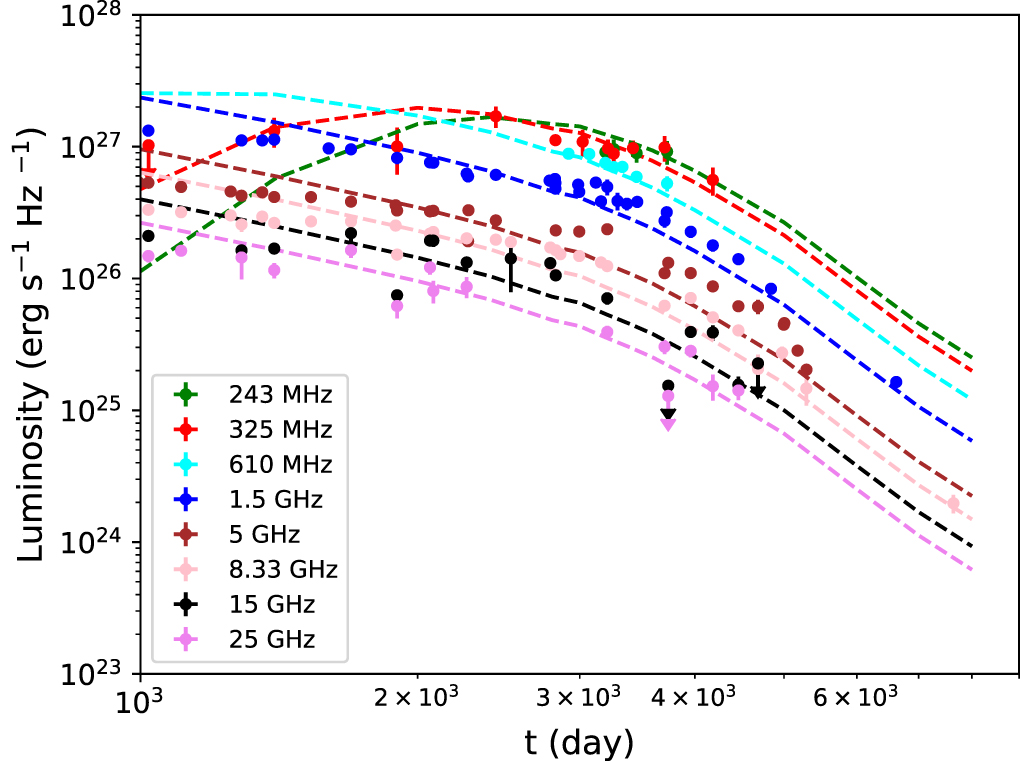}
\caption{{Radio} 
 light curves of SN 1993J after 1000 days from explosion. Filled circles represent the detections, while the arrows show the 3$\sigma$ upper limits. The~radio fluxes show a sudden downturn beyond around 3000 days. Before~this decrease, the~ambient medium can be characterized as $\Mdot/v_w = 4 \times 10^{-5} ~ \msunyr/10 ~{\rm km ~ s^{-1}}$. See text for details. Figure reproduced from \citep{kundu19} with~permission.}
\label{ek_sn1993j}
\end{figure}

\begin{figure}[H]
\includegraphics[width=13.7cm,angle=0]{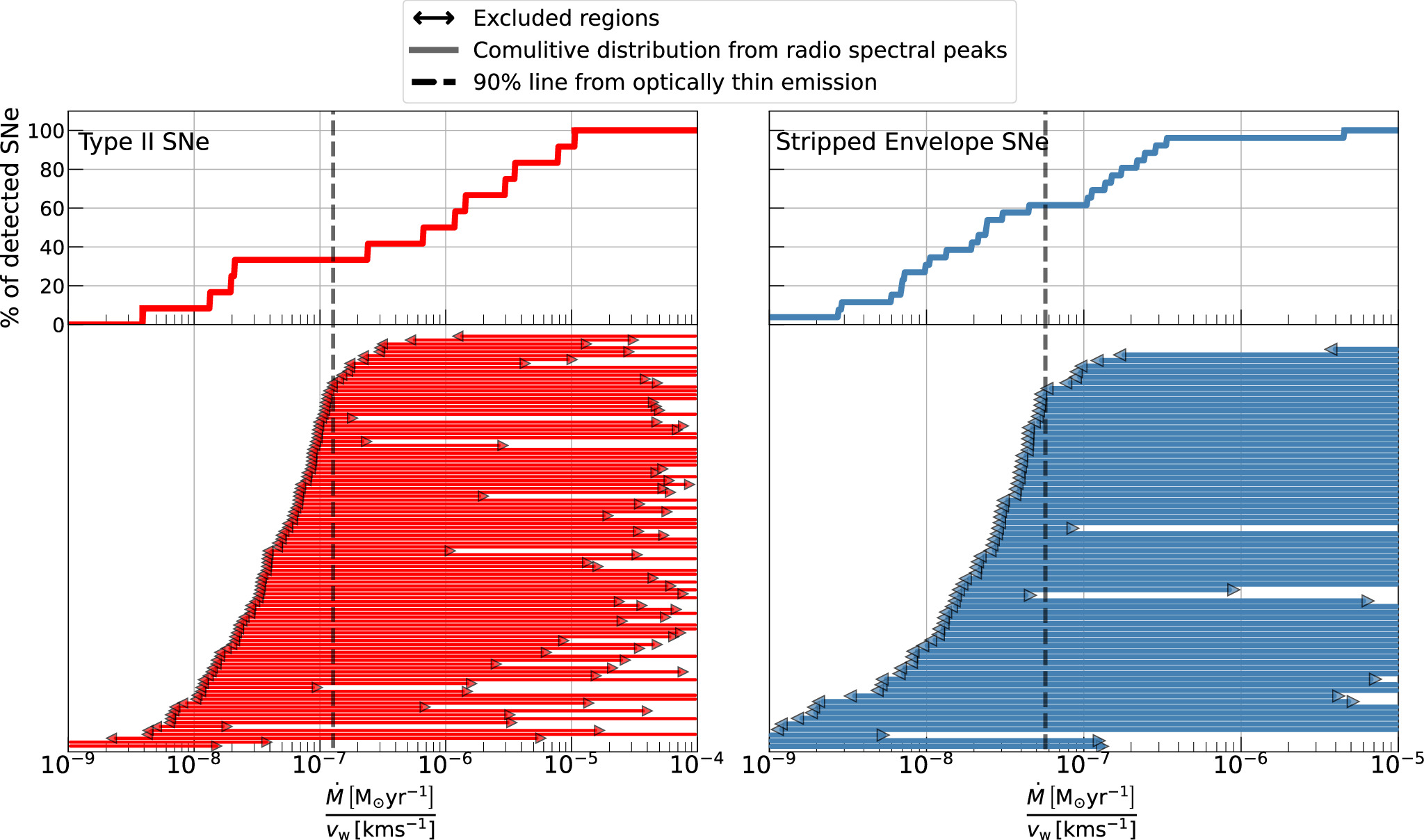}
\caption{{The} 
 phase space for $\mdot (\msunyr)/v_w (\kms)$ for radio-detected and non-detected SNe for Type II SNe  (\textbf{left} panel) and stripped-envelope SNe (\textbf{right} panel). Figure reproduced from \citep{Sfaradi25} with permission.  At~the top of the figure, the `Comulitive distribution' should be read as `Cumulative distribution'.
}
\label{mass_loss_phase_space}
\end{figure}

\newpage
The radio light curves of SN 1993J have shown a sudden decrease in fluxes beyond around 3000 days after the explosion, as~shown in Figure~\ref{ek_sn1993j}. Using the ejecta profile from STELLA (see left panel of Figure~\ref{den_prof}) and performing the hydrodynamical simulation of SN-CSM interaction, Kundu~et~al. (2019) \citep{kundu19} have concluded that this downturn of the radio fluxes could be associated with a much lower mass-loss rate beyond around 6600 $(v_w/10 ~{\rm \kms} )$ years prior to the explosion. Beyond~the main-sequence phase of the primary stage
, the~progenitor started to lose matter at a much higher rate, which caused a dense CSM around the SN. A~similar decrease in X-rays is also observed around a similar time. By analyzing both radio and X-ray evolutions of SN 1993, it was found that a wind-like ambient medium with $\Mdot/v_w = 4 \times 10^{-5} ~ \msunyr/10 ~{\rm km ~ s^{-1}}$ exists up to a radius of $2 \times 10^{17}$ ($\equiv$$R_{chng}$) cm, which represents the evolution of the SN up to around 3000 days after the explosion. Beyond~this, the~density of the CSM decreases rapidly. To~reproduce the evolution of radio and X-ray fluxes after 1000 days, the~CSM is proposed to have a density profile of {\citep{kundu17}}
\beq
\rho_{CSM} (r) = \rho_{CSM}^1(r) ~ \bigg [0.05 + \frac{0.95}{1+ \big(\frac{r}{4\times 10^{17}} \big)^4}  \bigg] ~ {\rm g/cm^3} 
\label{eq_rho_csm}
\eeq
where  $\rho_{CSM}^1(r) = \Mdot/(4 \pi v_w r^{2})$, with  $\Mdot/v_w = 4 \times 10^{-5} ~ \msunyr/10 ~{\rm km ~ s^{-1}}$. As~the density of the CSM decreases beyond $R_{chng}$, when modeling the radio light curves, as~displayed by the dashed lines in Figure~\ref{ek_sn1993j}, the~free--free optical depth will have the form  $\tau_{ff} = \frac{1}{3} r_s ~ \alpha_{ff}(r_s)~ \big(1 - f^3_{R_{chng}} (1 - f^2_{rate})  \big) $, where $f_{R_{chng}} = r/R_{chng}$ and $f_{rate} = \frac{\Mdot^1/v_w^1}{\Mdot/v_w}$, respectively. At~$r > R_{chng}$, the~medium is characterized by ${\Mdot^1/v_w^1}$. The~recent phase space analysis of $\mdot (\msunyr)/v_w (\kms)$ by Sfaradi~et~al. (2025) \citep{Sfaradi25}, using radio-detected and non-detected Type II and stripped-envelope SNe, has ruled out a mass-loss rate that lies in the range of $10^{-4}$ $\msunyr$ to $2 \times 10^{-6}$ $\msunyr$ for wind velocity $v_w = 10 ~ \kms$, as~shown in Figure~\ref{mass_loss_phase_space}.  

\begin{figure}[H]
\includegraphics[width=13.5cm,angle=0]{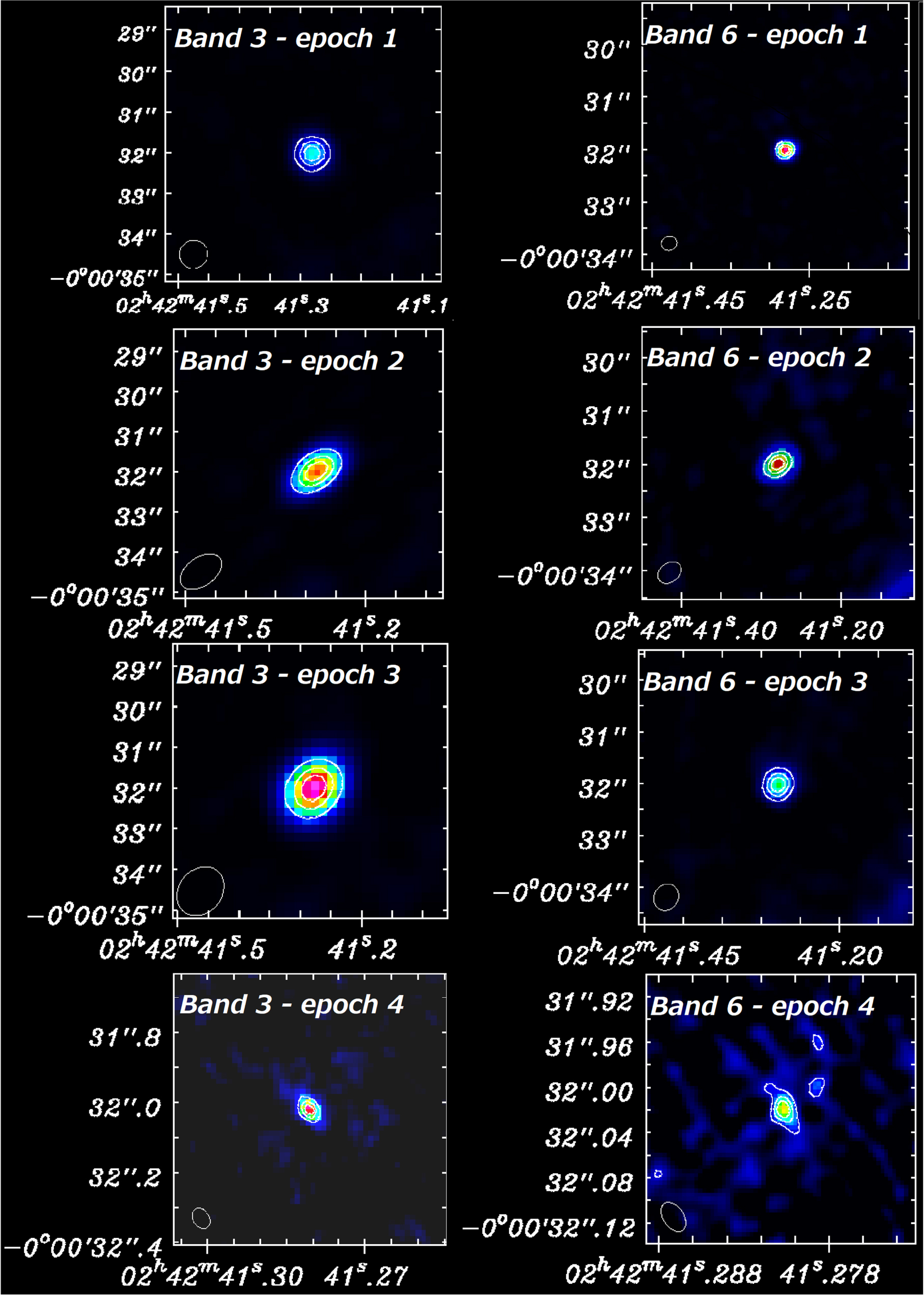}
\caption{{Radio} 
 images of SN 2018ivc at 100 and 250 GHz, using ALMA,~4.1, 7.1, 17.1, and 198.6~days after the explosion. Band 3 corresponds to the 100 GHz observations, and Band 6 represents the observations at 250 GHz. The~contours here represent 35\%, 60\% and 80\% of the peak flux density. At~100 GHz, For the first three epochs, the~color is normalized by the flux density in the range 0 to 10 mJy, and~for the first three 250 GHz observations, it is in the range of 0 to 5~mJy. For~the 4th epoch of observations,~color normalization is performed in ranges of 0 to 0.4 mJy for Band 3 and~0 to 0.2 mJy for Band 6. Figure reproduced from \citep{maeda23} with~permission.}
\label{radio_2018ivc_images}
\end{figure}

\newpage
 A peculiar Type IIL/IIb explosion is SN 2018ivc, for~which multiwavelength emissions are found to be primarily powered by SN-CSM interactions \citep{Bostroem20}. The~radio images of this SN at 100 (Band 3) and 250 GHz (Band 6), using ALMA are shown in Figure~\ref{radio_2018ivc_images} for four~different epochs since the explosion. As~displayed, from~epochs 1 to 3, the SN gets brighter at 100~GHz, whereas it maintains almost similar brightness in the first two epochs at 250~GHz. In~the last epoch, the~SN has reduced flux densities at both frequencies. It is concluded from the radio study that the mass-loss rate of the pre-SN star was almost five~times higher than that of SN 1993J in the last 200 years of its evolution \citep{maeda23}. The~history of mass loss of this SN suggests that it is a transitional event from a Type II/IIL to a Type~IIb/Ib/Ic~event.

Furthermore, using Markov Chain Monte Carlo analysis on 32 core-collapse SNe for which clear peaks were detected in their light curves, Matsuoka~et~al. (2025) \citep{Matsuoka25} predicted that the mass-loss rate of stripped-envelope SNe could be an order of magnitude higher than that of Type II SNe. Although~this investigation did not rule out the possibility of energy equipartition between the magnetic field and electrons, the~authors found both $\epsilon_e, \epsilon_{\rm B} < 10^{-2}$.

\par
\begin{figure}[H]
\includegraphics[width=12cm,angle=0]{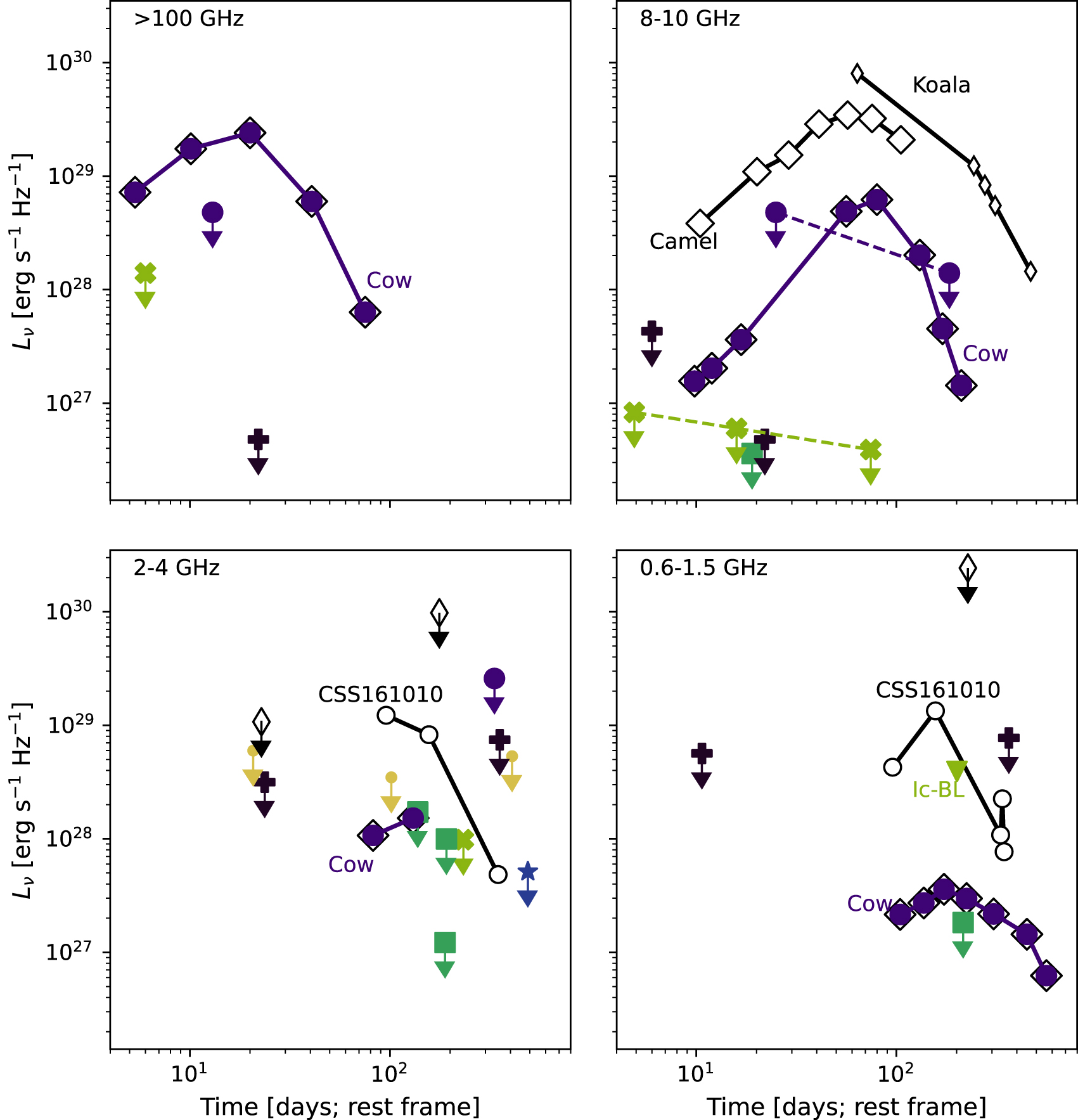}
\caption{{The} 
 radio and millimeter observations of Zwicky Transient Facility (ZTF)-detected FBOTs. Here, the big purple circles, black pluses, green crosses, squares, stars, diamonds, and small gold circles represent  IIn/Ibn, Ibn, Ic-BL, IIb, Ib, featureless, and unknown types of events. Among~the ZTF-detected FBOTs, three are categorized as Type Ibn, one as Type IIn, and one as Type IIn/Ibn explosion. As~seen, these events are much less radio-bright compared to radio-loud FBOTs such as Cow/AT2018cow, Koala/AT2018lug \citep{Ho19,Margutti19,Bietenholz20, Coppejans20,nayana21,Ho23}, CSS161010 \citep{Coppejans20}, and Camel/AT 2020xnd \citep{Ho22}. The~Ic-BL SN included in this comparison is the rapidly rising stripped-envelope event SN2018gep~\citep{Ho19a}. Figure reproduced from \citep{Ho23} with~permission.}

\label{FBOT_radio_mm}
\end{figure}

 
Apart from these, in~the past decade, with~the large number of multiwavelength surveys, a~different class of core-collapse SNe has been discovered; it has shown rapid evolution at optical wavelengths compared with other traditional SNe and was therefore named fast blue optical transients (FBOTs; \citep{Drout14,Pursiainen18,Rivera18,Prentice18,Margutti19,Ho19,Perley19}). The~radio and millimeter observations of Zwicky Transient Facility (ZTF) detected FBOTs, as~shown in Figure~\ref{FBOT_radio_mm}. Among~the ZTF-detected FBOTs, three were categorized as Type Ibn, one as Type IIn, and~one as Type IIn/Ibn explosion. As~seen, these events are much less radio-bright compared with radio-loud FBOTs such as Cow/AT2018cow, Koala/AT2018lug \citep{Ho19,Margutti19,Bietenholz20, Coppejans20,nayana21,Ho23}, CSS161010 \citep{Coppejans20}, and~Camel/AT 2020xnd \citep{Ho22}. The~Ic-BL SN included in this comparison is the rapidly rising stripped-envelope event SN2018gep \citep{Ho19a}.

\subsection{Thermonuclear SNe (Type Ia Explosions)}
\label{rad_sne_ia}
 Thermonuclear explosions of WDs can occur mainly through two channels, the~SD and DD channels. As~explained in Section~\ref{csm}, for~both scenarios, the~density profiles of the circumbinary media are different. As~a result, the~evolution of the radio fluxes differs for the SD and DD channels. This provides stringent constraints on the pre-SN systems, which are very poorly understood for Type Ia explosions. However, despite several attempts made to detect these SNe at radio wavelengths, we were unsuccessful in detecting radio emission from all these events except from one very recent Type Ia SN, SN~2020eyj.  
 
 \par
\begin{figure}[H]
\includegraphics[width=0.98\textwidth]{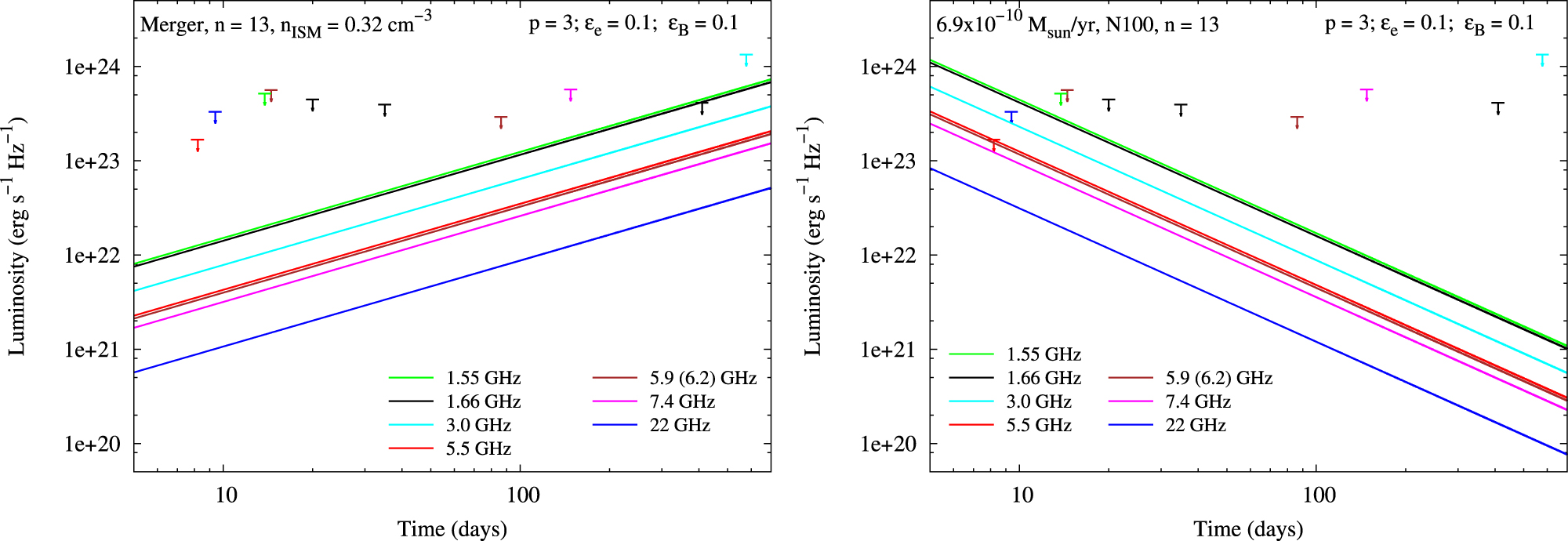}
\caption{{Radio} 
 light curves of SN 2014J when it plows through a constant-density medium (\textbf{left} panel) and a wind-like medium (\textbf{right} panel). The arrows represent the 3$\sigma$ upper limit on radio fluxes at different frequencies. For~$\epsilon_e = \epsilon_B = 0.1$, $n=13$, and~$p = 3$, late epoch observations at 1.66 GHz limit $n_{ISM} < 0.32 ~ \rm cm^{-3}$. For~the same parameters, the~observation performed around 8 days after the explosion at 5.5 GHz limits $\Mdot < 6.9 \times 10^{-10}$ $\msunyr$  for $v_w~=~100~\kms$.  Figure reproduced from \citep{kundu17} with~permission.}
\label{esha_ia_14J}
\end{figure}

 
 The nearest Type Ia explosion in the last few decades is SN 2014J \citep{goobar14} at a distance of around 3.4 Mpc, which gave us an enormous opportunity to study the circumbinary medium and thus explore the progenitor system of this event. To~achieve these goals, this SN was observed at radio wavelengths from very early times, around 8 days after the explosion, to~around 1.5 years \citep{pt14, kundu17}. None of these observations yielded detections. However, the~upper limits obtained at different radio frequencies provide stringent constraints on the density of the CSM for both SD and DD scenarios. In~Figure~\ref{esha_ia_14J}, the~radio light curves of SN 2014J with the upper limits and the models, in~solid lines, for~the DD (left panel) and SD (right panel) channels are shown. Assuming equipartition of energy between electron and magnetic fields with 10\% of the bulk kinetic energy channeling to each of them, i.e.,~$\epsilon_e = \epsilon_B = 0.1$ and~$p = 3$, the~observation performed around 8 days after the explosion at 5.5 GHz limits $\Mdot < 6.9 \times 10^{-10}$ $\msunyr$  for $v_w = 100 ~ \kms$ (see the right panel of Figure~\ref{esha_ia_14J}). For~the same shock parameters, late-epoch observations at 1.66 GHz limit $n_{ISM} < 0.32 ~ \rm cm^{-3}$ (see the left panel of Figure~\ref{esha_ia_14J}). Both provide the deepest limits on the CSM density for any Type Ia SNe to date, ruling out a significant number of pre-SN systems being the progenitors of SN 2014J. The~radio non-detections and the modeling suggested that this SN could be the result of the merger of two WDs, which justified the tenuous medium around the event. Although~these findings left little room for the SD scenario, the~spin-up/-down channel and the recurrent novae system can cause very-low-density media to exist around the pre-SN system (see \citep{kundu17} for details).                    

\begin{figure}[H]
\includegraphics[width=0.9\textwidth]{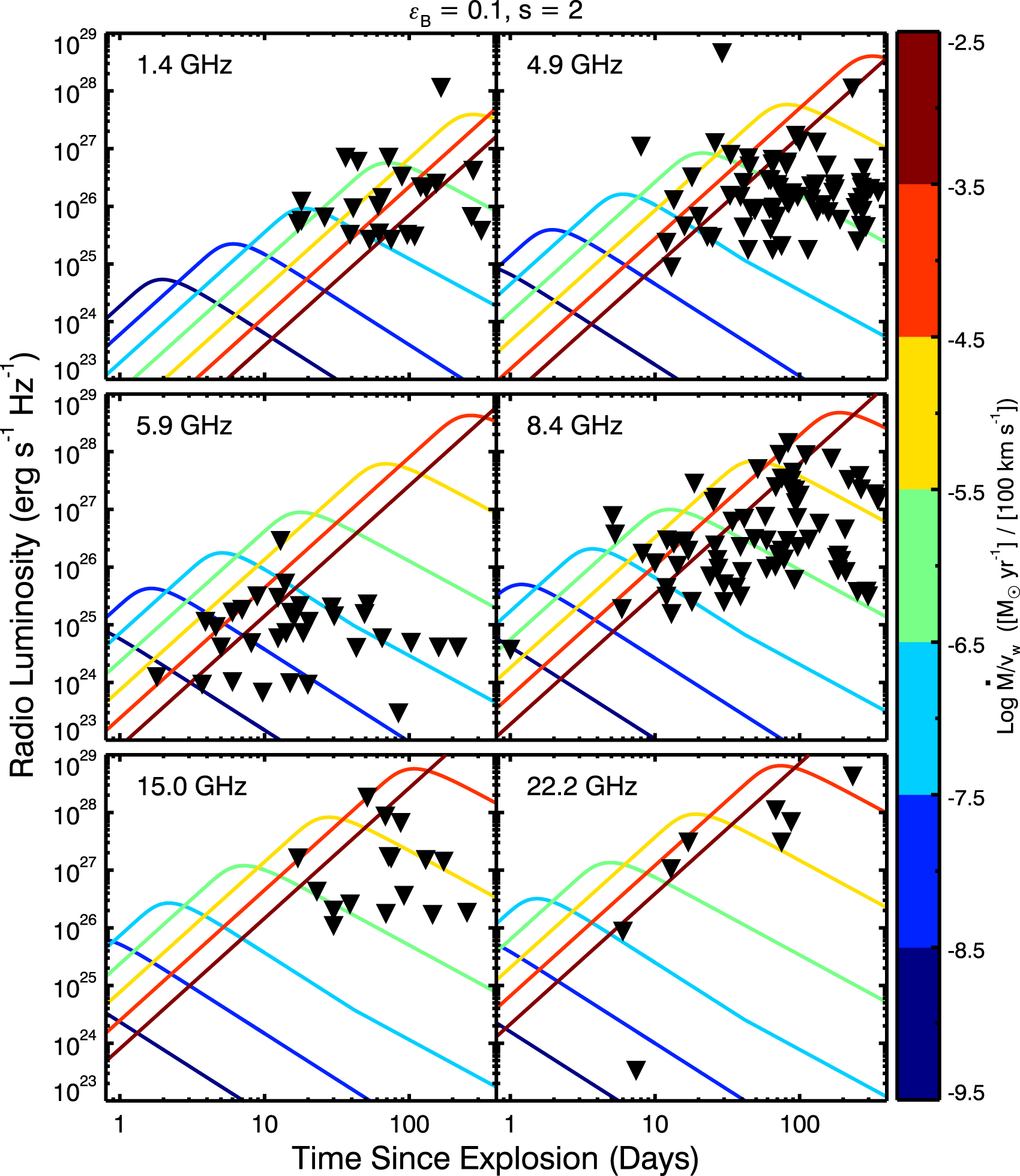}
\caption{{Radio} 
 light curves of various Type Ia SNe at different radio frequencies for a wind-like medium (i.e., $s = 2$) and $\epsilon_B = 0.1$. The~inverted triangles are three sigma upper limits on radio emissions from different Type Ia SNe. The~color bar on the right displays the different mass-loss rates of the pre-SN system in different colors for $v_w=$ 100 \kms. Figure reproduced from \citep{laura16} with~permission.}
\label{Laura_rad_Ia_sn}
\end{figure}

To explore the progenitors of these events, radio emission from an ensemble of Type Ia explosions was examined \citep{laura16, peter20}. The~radio light curves of different Ia SNe at various radio frequencies for a wind-like medium (i.e., $s = 2$) and $\epsilon_B = 0.1$ are shown in Figure~\ref{Laura_rad_Ia_sn}. Assuming $\epsilon_e = \epsilon_B = 0.1$, the~comparisons between the radio model and the observed $3\sigma$ upper limits constrain $\Mdot$ in the range of $10^{-3}$--$10^{-9}$ $\msunyr$ for $v_w = 100 ~\kms$.    

\begin{figure}[H]
\includegraphics[width=0.88\textwidth]{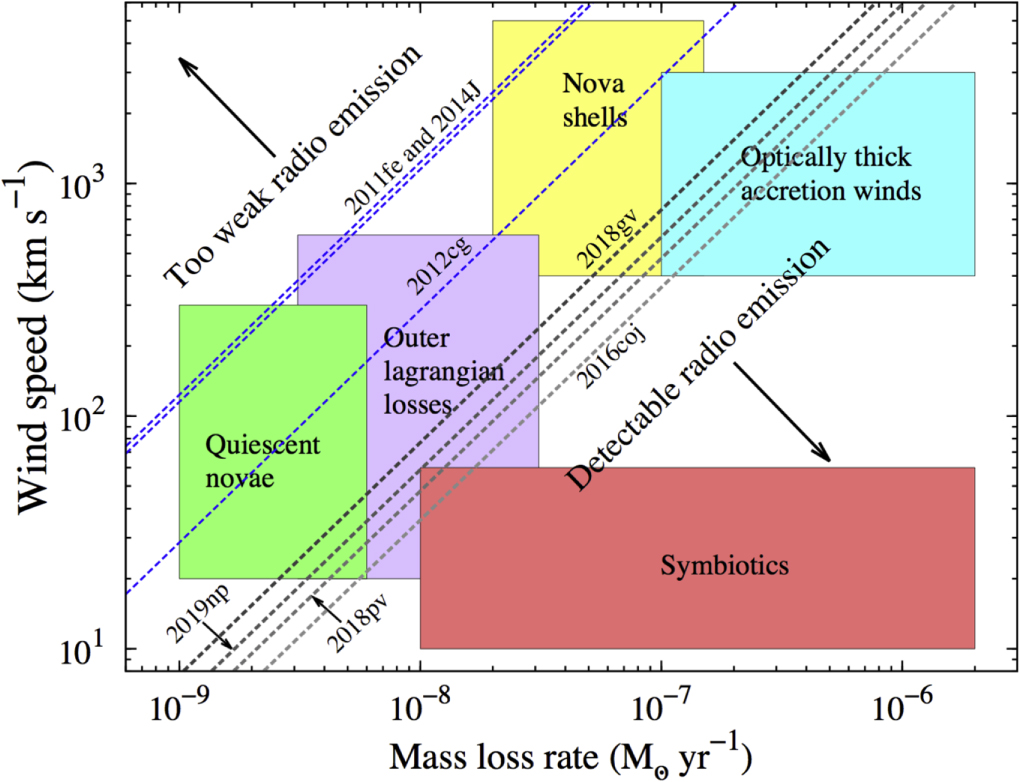}
\caption{For the SD scenario of thermonuclear explosion, the~constraints on the parameter space. For~low value of $\Mdot$ and high $v_w$, the~radio emission would be feeble, which is labeled as the Too weak radio emission regime in the figure. This scenario is true for SN 2014J (see text for details), which rules out a significant number of pre-SN systems for this SN. A~similar conclusion can be drawn for SN 2011fe, which is another nearby Type Ia explosion. For~other thermonuclear explosions, the~possible progenitor systems within the SD scenario are also displayed. Figure reproduced from \citep{peter20} with~permission.}
\label{peter_SD_constraint_20}
\end{figure}

Constraints on the mass-loss rates for different Type Ia explosions obtained from these analyzes allow us to rule out and limit the possible progenitor systems for the SD scenario. As~shown in Figure~\ref{peter_SD_constraint_20}, in~the case of SN 2014J, the~modeling of radio non-detections allows only a small parameter space to be consistent with the observations, favoring the DD scenario for this event. Another nearby Type Ia explosion is SN 2011fe, for~which a similar conclusion is drawn using rigorous modeling and radio observations for up to 4 years since the explosion (see \citep{kundu17} for details). For~other thermonuclear explosions, the~possible progenitor systems within the SD scenario are also displayed here. In~the case of the presence of a symbiotic companion, like a red giant star, the~radio emissions are expected to be much brighter. Since, for these SNe, no radio emission is detected, the~possibility of the presence of a symbiotic star as a secondary star is ruled out for most of the progenitor~systems.        

\begin{figure}[H]
\includegraphics[width=0.88\textwidth]{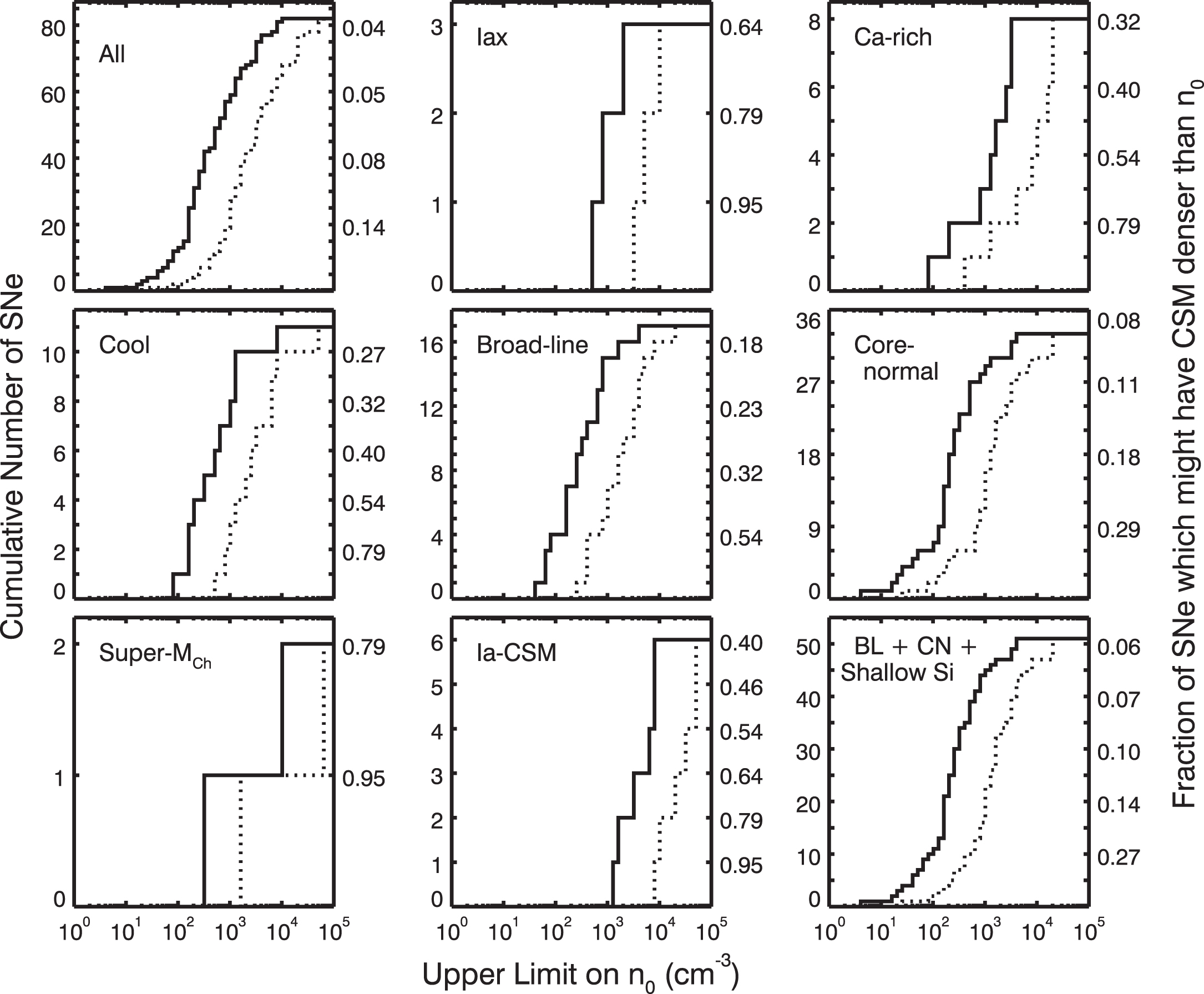}
\caption{{Upper} 
 limits on the CSM particle density from radio observations for the DD scenario for different subgroups (for details about the subgroups, see \citep{laura16} and references therein). The~solid cumulative histograms give the limits for $\epsilon_B = 0.1$, and the dotted histogram is that for $\epsilon_B = 0.01$. Figure reproduced from \citep{laura16} with~permission.}
\label{laura_nISM}
\end{figure}

Categorizing around 80 Type Ia SNe according to their rise time in the B-band into~eight~different subgroups, Chomiuk~et~al. 2016 \citep{laura16} estimated the upper limits of the CSM particle density from radio observations for the DD scenario, which is shown in Figure~\ref{laura_nISM}. The~solid histograms give the limits for $\epsilon_B = 0.1$ and the dotted histogram is that for $\epsilon_B = 0.01$. As~demonstrated in Figure~\ref{laura_nISM}, for~some Type Ia events
, the~density can be as high as $\sim$$10^{5}$ cm$^{-3}$, although~for most of them, the ambient media have low~densities. 

\begin{figure}[H]
\includegraphics[width=0.6\textwidth]{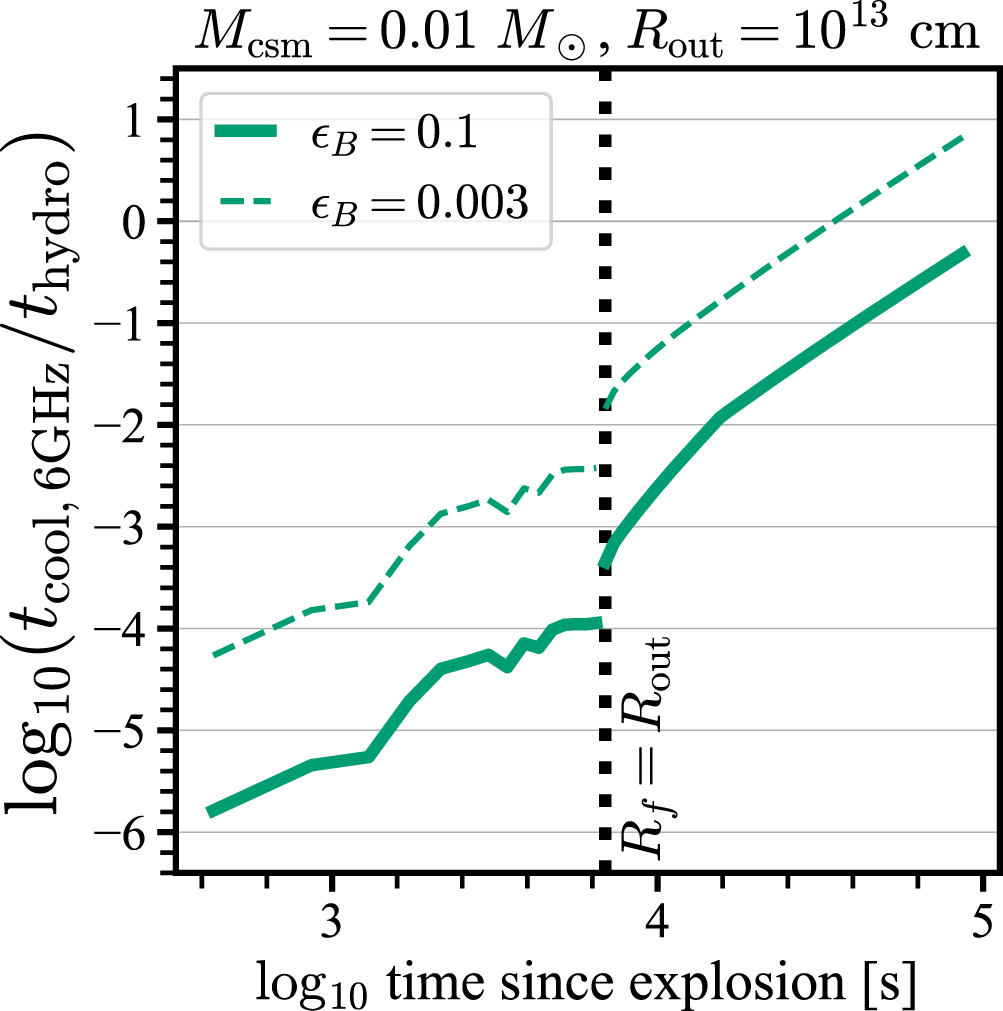}
\caption{Cooling efficiency of relativistic electrons responsible for radio synchrotron radiation behind~the shock front as~a function of time for different magnetic field strengths. Figure reproduced from \citep{Harris23} with permission. The~black dotted vertical line represents the time at which the compact shell is overtaken by the shock. 
}
\label{t_cool_rel_elec}
\end{figure}

To further investigate the non-detections of SNe of Type Ia at radio wavelengths, the~efficiency of the cooling of the relativistic electrons responsible for radio synchrotron radiation behind~the shock front as~a function of time for different magnetic field strengths is examined by Harris~et~al. (2023) \citep{Harris23} using the SuperNova Explosion Code \citep{Morozova15,piro16}. They concluded that after interaction, the~relativistic electrons cool at a significantly faster pace, as shown in Figure~\ref{t_cool_rel_elec}.

\begin{figure}[H]
\includegraphics[width=0.8\textwidth]{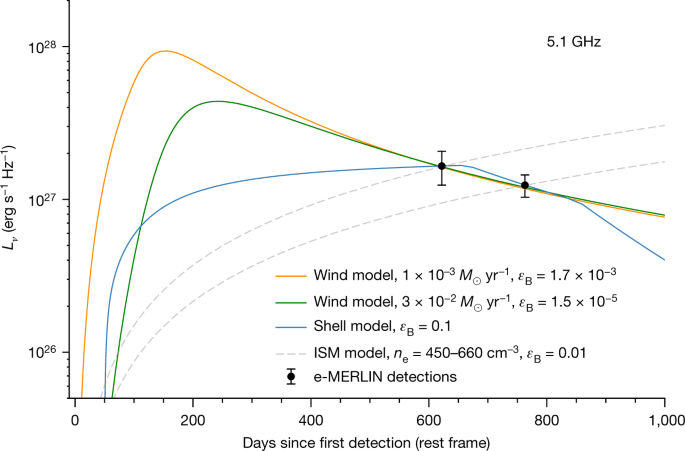}
\caption{{SN} 
 2020eyj radio light curves at 5.1 GHz. This is the only Type Ia SN that is detected at radio wavelengths, shown here by the filled circles (the observations were conducted with the e-MERLIN telescope). For~both wind-like and constant-density medium, the~radio emissions are modeled here to get the mass-loss rate of the progenitor systems or the ISM density for different values of $\epsilon_B$. Figure reproduced from \citep{kool23} with~permission.}
\label{2020eyj_rad}
\end{figure}

Until now, the~single Type Ia event detected at radio wavelength is SN 2020eyj \citep{kool23}. The~radio observations carried out around 605 and 741 days after the explosion of this SN yielded the detection at 5.1 GHz. The~radio detections along with the model predictions are shown in Figure~\ref{2020eyj_rad}, suggesting an SD explosion channel for this event. The~detections are consistent for a wind model (i.e., $s=2$) with $\Mdot$ between  $3 \times 10^{-2} ~ \msunyr$ and \mbox{$1 \times 10^{-3} ~ \msunyr$} for a $v_w = 1000 ~{\rm km ~ s^{-1}}$ when $\epsilon_B$ varies between $10^{-5}$ and $10^{-3}$. The~detections are also consistent with a shell model, where the CSM is concentrated in a shell. If~this SN was due to the merger of double WDs, radio emission predicts that the ambient medium should then have a much higher density, with $n_{ISM} \sim 450$--$600$ $\rm cm^{-3}$.

\unskip

\unskip

\begin{figure}[H]
\includegraphics[width=0.7\textwidth]{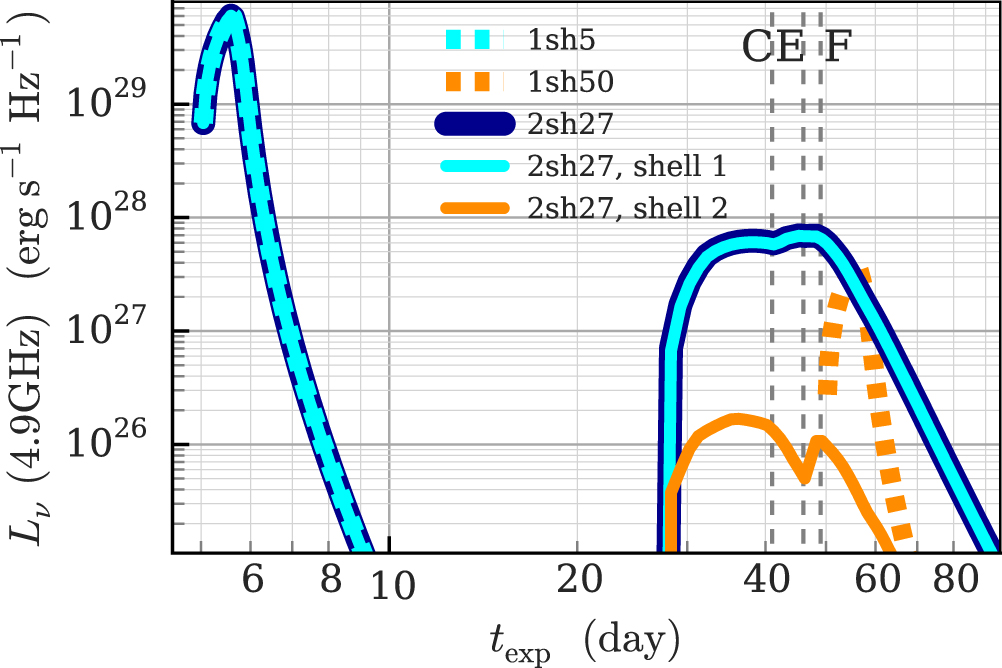}
\caption{ 
{Radio} 
 emissions from dense CSM shells, created due to recurrent novae. The single-shell models are represented by thick dashed lines. Figure reproduced from \citep{Harris16} with permission.
}
\label{Harries_2016}
\end{figure}

As several observation campaigns aimed at detecting radio emission from Type Ia events resulted in non-detections \citep{Panagia06,Horesh12,Chomiuk12,Margutti12,Silverman13,pt14,kundu17,peter20,Harris23}, nova outbursts could be a common cause that occurs in most Ia systems before explosions. These outbursts sweep the CSM into dense shells away from the pre-SN system and~therefore create a tenuous medium close to the progenitor site \citep{Wood-Vasey06,Patat11,Moore12}. The~radio emissions from these dense CSM shells were investigated by Harries~et~al. in 2016 \citep{Harris16}, as~shown in Figure~\ref{Harries_2016}, who demonstrated that the radio emission expected from the two-shell model (2sh27) is drastically different from that of the one-shell model (1sh5, 1sh50). Additionally, there are episodes of no emission from the low-density media that exist between the~shells.

\section{Discussion}
\label{discusn}
The study of radio emissions from SNe is an excellent probe for gaining information about progenitor stars through their mass-loss history before the explosion. Several core-collapse SNe have shown excessive mass losses just before collapse, which caused enhanced radio emission from these events. Due to the gaps in our understanding of what happens just prior to and during the explosion and the lack of direct observational proof of stars' behavior during these phases, the~study of radio emission provides compelling information that allows us to explore the final stages of stars' life leading to the explosions, which otherwise would not be possible to~investigate. 

\par 
Apart from the above, resolving SNe at radio wavelengths gives the maximum extent of the event and enables us to get information about possible asphericity in the explosions
. As~seen in the case of SN 1993J, the~radio emissions trace a nearly spherical shell. Moreover, also for~some SN remnants, similar spherical shell structures have been observed \citep{miroslav25}. The~best way to understand the expansion of the shocks,~hence the~SN, is through VLBI/VLBA observations because of their imaging capability at the milliarcsecond level, as~described in the previous section. While a microarcsecond level of accuracy can be obtained using the Event Horizon Telescope, it can be achieved mainly for objects with higher brightness temperatures and frequencies higher than around 250~GHz.
As mentioned earlier, a~milliarcsecond level of accuracy is essential to understanding the GRB-SN connection and constrain the immediate surroundings of the explosions. Furthermore, a~couple of fast radio burst (FRBs) events are found to be associated with SN explosions~\citep{Bochenek20,chime20,kundu20,kundu22}. Since FRBs are mainly cosmological in origin, to~investigate the connection between SNe and FRBs, Long-Baseline Interferometric observations have been proven to be fruitful \citep{Chatterjee17,Ravi22,cassanelli24,Hewitt24}.

\par 
Type Ia explosions are considered standard candles in cosmology; however, the~progenitors of these explosions are poorly constrained. Although, to~date, we have only one thermonuclear SN detected at radio frequencies, most Type Ia SN remnants show detectable radio emissions (e.g., see \citep{Dubner15}). It is expected that a certain percentage of these explosions arise from the SD channels; however, the~radio non-detections point towards a significant number of these events being due to the merger of two WDs. On~the other hand, recurrent novae, which can cause a tenuous medium to exist around the pre-SN system, could be the reason for a significant number of these~non-detections. 

\par
Besides providing stringent constraints on pre-SN systems, rigorous investigations of radio emissions from SNe allow us to understand the microparameters of shock physics. SN shocks have high Mach numbers ($>$$1000$), making them computationally unfathomable because of current computational limitations. As~a result, the~best diagnostic tool for exploring shock physics is through the investigation of radio emissions that originate from the shock-accelerated particles in the presence of magnetic fields, which get amplified in shocked shells. Using the radio observations of the nearest-two Type Ia explosions, SN 2014J and SN 2011fe, Kundu~et~al. (2017) tried to constrained the~amplifications of magnetic fields in SN shocks \citep{kundu17}
. This investigation favors low amplification of magnetic fields in SN shocks with $\epsilon_B < 0.1$.

\section{Conclusions}
\label{conclu}
With numerous applications, SNe, whether detected or not detected at radio wavelengths, enrich our understanding of a wide variety of objects and phenomena in the universe, ranging from exploring different types of stars leading to explosions to shock physics and explosion mechanisms.  
Wide-field all-sky radio surveys, such as the VLASS, with~sky coverage above declination $> -40 ~\!\!^\circ$ (equivalent to a coverage of around {33,885} 
 ${\rm deg^2}$), in~the frequency window of 2--4 GHz \citep{Lacy20}, and~ASKAP VAST Pilot surveys, covering the southern hemisphere, have been successful in detecting the rebrightening of SNe at later stages of their lives \citep{Stroh21,Rose24}. These detections open new pathways for reconstructing the mass-loss history of progenitor stars much earlier than their catastrophic destruction occurred, which provides stringent constraints on the evolution of progenitor stars that are otherwise impossible to~retrieve. 

\par 
The current generation radio telescopes, such as the LOw-Frequency ARray (LOFAR)~\citep{vanHaarlem13}, the upgraded Giant Meterwave Radio telescope \citep{Swarup91,Ananthakrishnan95}, MeerKAT \citep{Jonas16}, the ASKAP, Parkes \citep{Robertson12}, the Green Bank Telescope \citep{Jewell04}, the VLA \citep{Thompson80,Napier83,Perley11}, the Australia Telescope Compact Array~\citep{Ricci04}, and the ALMA \citep{Wootten09}, along with VLBI/VLBA, have made it possible to observe SNe in regimes from as low as 100 MHz to a few hundred GHz.  {{This wide range of coverage, combined with improved resolution, enables us to explore the underlying physics of explosions in greater detail and address the open issues in this field. The~lack of detection of thermonuclear SNe at radio wavelengths raises significant doubts about the nature of their pre-SN systems. At~present, most predictions regarding the progenitors of Type Ia events are based on early-time radio observations. However, for~a clearer picture, periodic and systematic observations of these explosions at early and late times are crucial. In~the case of core-collapse SNe, although~around 30\% of the explosions are detected at radio frequencies, for~almost all (except~for a few) explosions, the~data are not well sampled, introducing larger uncertainties when determining the progenitor's mass-loss episodes at different stages. In~addition to these, other observational issues involve the non-detection of SNe as a result of the inadequate sensitivity of the~telescopes.} 
}

{{Another major issue regarding the interpretation of radio emissions arises because of superficial knowledge of shock physics. The~acceleration of charged particles and~amplifications of magnetic fields, which play a crucial role in estimating the intensity of radio emission in SN shocks, are poorly constrained.  Furthermore, there is ample evidence of asymmetric explosions and inhomogeneous matter distributions in ambient media. As~a result, assumptions of spherical shock structure and homogeneous CSM distribution, which are widely considered when estimating radio emissions from SNe, prevent us from determining radio emission properties with higher accuracy.}} From the point of view of cosmology, the~progenitors of Type Ia explosions are crucial to understanding and exploring the expanding universe to a greater extent. Nevertheless,~21 cm radio observations offer us the opportunity to delve into the universe up to a redshift of around 150, where the first structures of the universe and stars are expected to be created (for~details see, e.g., \citep{Pritchard12}). The~upcoming/next-generation radio telescopes, such as the Square Kilometre Array (SKA) \citep{Dewdney09}, the next-generation VLA (ngVLA) \citep{Francesco19}, and~LOFAR2.0 (\textls[-37]{{\url{https://www.lofar.eu/wp-content/uploads/2023/04/LOFAR2\_0\_White\_Paper\_v2023.1.pdf}, accessed on 30 May 2025),
}} \textls[-15]{are~expected to reshape SN radio astronomy by 
providing further insights into these~explosions.}

\par

\vspace{6pt}


\dataavailability{All the data used in this article are published; therefore, they can be acquired from the articles cited here.}


\acknowledgments {The author thanks the anonymous referees for their helpful comments.}

\conflictsofinterest{The author declares no conflicts of~interest.} 

\abbreviations{Abbreviations}{
The following abbreviations are used in this manuscript:
\\

\noindent 
\begin{tabular}{@{}ll}
SN & Supernova \\
SLSN & Superluminous Supernova \\
GRB & Gamma-Ray Burst \\
RT & Rayleigh--Taylor \\
CSM & Circumstellar medium \\
WD & White dwarf \\
SD & Single degenerate \\
DD & Double degenerate \\
SED & Spectral energy distribution \\
VLBI & Very-Large-Baseline Interferometry \\
VLBA &  Very-Long-Baseline Array \\
VLA & Very Large Array \\
VLASS & Very Large Array Sky Survey \\
VAST & Variable and Slow Transients \\
ASKAP & Australian Square Kilometre Array Pathfinder \\
ALMA & Atacama Large Millimeter/submillimeter Array \\
uGMRT & upgraded Giant Meterwave Radio telescope \\
LOFAR & LOw-Frequency ARray \\
ATCA & Australia Telescope Compact Array \\
SKA &  Square Kilometre Array \\

\end{tabular}
}

\appendixtitles{no} 

\begin{adjustwidth}{-\extralength}{0cm}

\reftitle{References}

\PublishersNote{}
\end{adjustwidth}
\end{document}